\documentclass[journal]{IEEEtran}
\IEEEoverridecommandlockouts
\ifCLASSINFOpdf
\else
\fi
\usepackage{url}

\usepackage{amsmath}
\usepackage{bbm}

\usepackage{graphicx}
\usepackage{booktabs}
\usepackage{multibib}
\newcites{app}{Appendix References}

\usepackage{siunitx}
\usepackage{multirow}
\usepackage[table]{xcolor}
\usepackage{amsthm}  

\usepackage{csquotes}
\usepackage{amsmath, amsthm, amssymb}
\usepackage{standalone}
\usepackage{myPlotStyle}
\usepackage[capitalise]{cleveref}

\usepackage{tikz}
\usetikzlibrary{positioning}

\pgfplotscreateplotcyclelist{ntkstyles}{
  solid,        ultra thick,   black\\%
  dashed,       ultra thick,   sRed\\%
  dash dot,     ultra thick,   sGreen\\%
  dotted,       ultra thick,   sBlue\\%
  dash pattern=on 3pt off 1pt on 1pt off 1pt, ultra thick, sOrange\\%
  dash pattern=on 1pt off 1pt, ultra thick,        sPurple\\%
  dash pattern=on 5pt off 1pt, ultra thick,        sBrown\\%
  dash pattern=on 5pt off 1pt on 1pt off 1pt, ultra thick, teal\\%
}

\usepackage{acronym}
\usepackage{bm} 
\usepackage{url}
\newacro{NN}[NN]{Neural Network}
\newacro{PINN}[PINN]{Physics-Informed Neural Network}
\newacro{PIML}[PIML]{Physics-Informed Machine Learning}
\newacro{KAN}[KAN]{Kolmogorov-Arnold Neural Network}
\newacro{NTK}[NTK]{Neural Tangent Kernel}
\newacro{MLP}[MLP]{Multi-Layer Perceptron}
\newacro{ML}[ML]{Machine Learning}
\newacro{SM}[SM]{Synchronous Machine}
\definecolor{cbBlue}{RGB}{0,114,178}
\definecolor{cbOrange}{RGB}{230,159,0}
\definecolor{cbGreen}{RGB}{0,158,115}
\definecolor{cbVermillion}{RGB}{213,94,0}
\definecolor{cbPurple}{RGB}{204,121,167}

\renewcommand{\baselinestretch}{0.98}

\begin{document}
\newcommand{\sansserifformat}[1]{\fontfamily{cmss}{ #1}}%
\setlength{\abovecaptionskip}{0pt} 
\setlength{\belowcaptionskip}{0pt}

\title{Tools to Explain Neural Networks for Power System Dynamics}
\author{
\IEEEauthorblockN{Petros Ellinas, Johanna Vorwerk, and Spyros Chatzivasileiadis}
\IEEEauthorblockA{
Department of Wind and Energy Systems, Technical University of Denmark, Lyngby, Denmark\\
Email: \{petrel, vorjo, spchatz\}@dtu.dk}
\thanks{This work is supported by the European Research Council (ERC) Starting Grant VeriPhIED, Grant Agreement No. 949899.}
}

\maketitle


\begin{abstract}
This paper presents, for the first time in power systems literature to our knowledge, analytical tools to explain the training performance of machine learning surrogate models for power system dynamics. Power system simulations are increasingly challenged by stiff and multi-timescale dynamics arising from converter-interfaced resources and fast control loops. Machine learning surrogates emerge as promising tools to handle this complexity and accelerate dynamic simulations. However, their performance remains difficult to interpret, which limits their adoption.
Building on the small-signal eigenvalue analysis in power systems, this paper uses the Neural Tangent Kernel (NTK) method. NTK delivers a modal interpretation of the learning performance, identifying error modes that decay rapidly versus others that converge slowly. This connection explains how physical stiffness and timescale separation in power system dynamic models appear as optimization stiffness during Neural Network (NN) training. Based on this analysis, we develop adaptive loss-weighting strategies to improve and explain why structure-aware neural architectures, such as ActNet, perform better than vanilla NNs. We assess the proposed approach on physics-informed machine learning surrogate models of \acp{SM} and power electronic converters. The methods introduced in this paper can deliver the necessary analytical tools to interpret and improve the performance of machine learning surrogates, paving the way for the systematic, physics-aware design of NN architectures and training strategies. By moving beyond trial-and-error development, these tools reveal training dynamics and failure modes, support more reliable design decisions, and strengthen confidence in machine-learning surrogates for engineering applications.
\end{abstract}

\IEEEpeerreviewmaketitle

\section{Introduction}
The rapid growth of renewable generation and the electrification of heating and transport have increased the need for fast and reliable dynamic analysis of large-scale nonlinear power systems. Operators must assess system security across many contingencies and operating points.
Each time-domain simulation, however, is computationally demanding because the governing
Differential-Algebraic Equations (DAEs) are nonlinear and multi-timescale: fast electromagnetic
and control modes coexist with slow electromechanical modes. While all have been present in power systems dominated by \acp{SM}, converter-interfaced generation convolutes timescale separation and requires considering multiscale DAEs in dynamic simulations \cite{lara, uroc}. This need to consider a wide variety of time
scales enhances stiffness, which forces explicit solvers to take very small time steps and
causes implicit solvers to face increasingly ill-conditioned algebraic systems. When these
simulations must be repeated thousands of times to cover the wide range of credible scenarios, classical methods quickly become too slow for real-time or near-real-time decision support.

Many acceleration strategies have been proposed, including model reduction, network equivalencing,
high-performance computing implementations, and specialized techniques to accelerate electromagnetic
transient (EMT) simulations~\cite{seepdupEMT}. More recently, \ac{ML} approaches have been
explored to approximate the dynamics of specific power-system components, either by directly approximating system trajectories~\cite{STIASNY2023109748,Ellinas2024}
or by learning surrogate models of the underlying dynamics~\cite{bossart2024accelerationpowerdynamicsimulations}.
These efforts broadly follow two complementary directions:
(i) stand-alone surrogates that directly map operating conditions and time
to system trajectories for rapid open-loop screening%
~\cite{STIASNY2023109748,Ellinas2024}, and
(ii) learned surrogates embedded within a numerical
DAE solver~\cite{bossart2024accelerationpowerdynamicsimulations}.
Both approaches exploit \ac{NN} interpolation and generalization capabilities across operating conditions, enabling extremely fast evaluation after training.

Along this line of research, \ac{PIML} approaches have emerged that improve reliability and data efficiency by incorporating the physical structure directly into training and penalizing violations of the governing DAEs and initial conditions at collocation points~\cite{misiris}. The \ac{ML} model maps initial states, control inputs, and time to system trajectories, while minimizing the DAE residual to constrain the surrogate to the physically admissible solution manifold. When used in an open-loop setting, e.g., in a single-machine infinite bus, this enables the evaluation of several time points in parallel, rather than sequential time stepping, which drastically increases computation speed.

Here, it is interesting to highlight the analogy between physics-based power-system dynamic models formulated as DAEs and surrogate models based on \acp{NN}. In the ML-based surrogate modeling of dynamical systems, NN training determines the learnable parameters, i.e. weights and biases, by solving an optimization problem so that the model reproduces the time evolution of the states exactly as described by a physics-based system of DAEs. In this sense, the role of the explicit time integration of the DAE system is replaced by an optimization-based training procedure. As a result, classical notions of numerical stability of the solver reappear as questions of convergence, conditioning, and robustness of the underlying optimization process.

Despite their advantages, ML surrogates are still treated with caution in power-system applications because their internal behavior is often opaque, and it is unclear how design choices and training objectives affect accuracy and reliability. In safety-critical contexts, this opacity is also increasingly at odds with emerging expectations for transparency and auditability~\cite{europaRegulation20241689}. In this work, we view transparency not only as the interpretability of model internals but also as the ability to predict and explain reliability and failure modes, as we usually do from the structure of the underlying DAEs. This motivates a structural question: \emph{How does physical stiffness in the governing DAEs affect the numerical stability and reliability of the NN training optimization problem, and how can its failure modes be predicted?} In this paper, we study \ac{ML} surrogates for power-system components and address four central questions: (i) How do different neural architectures learn stiff and multi-scale dynamics? (ii) How do modeling choices affect accuracy and reliability? (iii) (iii) how reliability and failure modes, meaning when and why training becomes inaccurate or fails to converge, can be predicted and explained? and (iv) how these choices influence generalization beyond the training conditions? We treat the surrogate architecture, the choice of activation functions, and the design of the loss function as deliberate modeling choices meant to adapt to the mathematical structure of the governing DAEs, rather than heuristic design decisions. To answer these questions, we focus on individual components, both because
much of the recent literature studies component-level \ac{ML} surrogates~\cite{Ellinas2024} and because low-order models keep the NTK analysis interpretable, allowing the underlying mechanisms and failure modes to be isolated. The proposed analysis is not restricted to low-order systems and can also be applied to \ac{ML} surrogates of higher-order dynamic models.

To study how NN architectures interact with stiffness and the multi-timescale behavior of power system dynamics, we use the \ac{NTK} method \cite{jacot2020neuraltangentkernelconvergence} to analyze the training dynamics. More generally, \ac{NTK} provides a tool for analyzing the convergence of individual objectives in multi-objective NN training, where different loss components may drive parameter updates with different strengths and converge at different rates. In the \ac{PIML} surrogate setting considered here, these loss components correspond to the initial-condition and physics-residual terms. Analogous to eigenvalues in small-signal stability analysis, the \ac{NTK} provides a spectral decomposition of training dynamics: \ac{NTK} eigenvalues determine how different loss-function error modes decay, just as Jacobian eigenvalues characterize the post-disturbance evolution and decay of physical modes. This connects physical stiffness in the DAEs with optimization stiffness in training and guides the use of adaptive loss-weighting strategies that rebalance modes and improve convergence in mildly stiff regimes.
The main contributions of this work include:
\begin{itemize}
\item \textbf{A diagnostic tool for multi-objective learning.}
Using the \ac{NTK} method, we propose a comprehensive analysis tool for multi-objective training, including \ac{PIML} models. We present an analogy to small-signal stability analysis to provide an intuitive interpretation of the training dynamics. It identifies when training is well-posed or ill-conditioned by explaining how different loss components and embedded physical properties affect learnability and convergence. Furthermore, the NTK analysis guides the use of adaptive loss weighting to promote balanced convergence across loss terms.

\item 
We compare \ac{NN} architectures for stiff and multi-timescale dynamics, including ActNet. Based on observed performance and our NTK analysis, we provide practical guidelines for selecting architectures and activation functions according to the structural properties of the approximated dynamics.

\item 
We evaluate the proposed methods on \ac{SM} and inverter models (ranging from 2nd order to 11th order models), highlighting both their strengths and limitations across dynamical regimes.
\end{itemize}

Sections~\ref{sec:DAE}--\ref{sec:casestudies_results} present the DAE formulation, the PIML surrogate setup, the structure-aware neural architectures, the NTK training analysis, and the case-study results; the paper closes with conclusions on trustworthy surrogate modeling for power-system dynamics.

\section{Differential–Algebraic Systems for Power Systems}
\label{sec:DAE}

DAEs model power-system dynamics, which combine differential equations with algebraic constraints:
\begin{equation} \label{eq:general_dae}
\mathbf{M}\,\frac{d\mathbf{x}(t)}{dt} - f\bigl(\mathbf{x}(t),\mathbf{u}(t),t\bigr)=0,
\qquad \mathbf{x}(t_0)=\mathbf{x}_0 .
\end{equation}
Here $\mathbf{x}(t)\in\mathbb{R}^n$ collects both dynamic states, e.g., generator rotor angles, speeds, and controller states, and algebraic
variables, e.g., bus voltages. The input $\mathbf{u}(t)$
represents disturbances or control actions. The mass matrix $\mathbf{M}$
contains ones on differential rows and zeros on algebraic rows.
Rows with $M_{ii}=0$ impose instantaneous algebraic constraints,
such as network power balance, which are enforced at each time step
using a root-finding method, typically Newton–Raphson.

To simulate the DAEs, time-stepping schemes advance the solution either explicitly or implicitly. 
Explicit methods compute $\mathbf{x}_{k+1}$ directly from known quantities and are inexpensive per step, but require small time steps in the presence of fast dynamics. 
Implicit methods allow larger time steps and are therefore common in transient-stability programs, but require solving a nonlinear system
\(F(\mathbf{x}_{k},\mathbf{x}_{k+1})=0\) \cite{hairer1996solving}.

Power-system models are stiff because fast electrical or control states and slower electromechanical states are coupled within the same DAE system, even though their time scales may differ by several orders of magnitude. As a result, explicit solvers must use time steps small enough to remain stable for the fastest dynamics, even when the main quantities of interest evolve much more slowly. Implicit solvers alleviate this restriction, but at the cost of solving more challenging nonlinear systems. A more detailed discussion of stiffness and numerical-solver stability is provided in the companion document~\cite{githubGitHubElpetros99tools_explaining_power}. These challenges motivate surrogate approaches that approximate system evolution without a full nonlinear solve at every step.

\section{PIML for Differential--Algebraic Systems}
\label{subsec:PIML}

This section introduces the neural solution operator, its physics-informed
training objective, and the structure-aware architectures considered in
this work.

\subsection{Neural Networks}
\label{sec:mathfoundation}

\acp{NN} are a fundamental class of ML models composed of $L$ layers of
interconnected neurons. Formally, a \ac{NN} defines a parametric mapping
$\mathrm{NN}_{\boldsymbol{w}} : \mathbb{R}^{n_0} \to \mathbb{R}^{n_L}$,
which transforms an input vector $\mathbf{z} \in \mathbb{R}^{n_0}$ into an output
$\mathbf{y} \in \mathbb{R}^{n_L}$ through a sequence of layer-wise transformations
$\boldsymbol{\Phi}_l$:
\begin{equation}
\mathrm{NN}_{\boldsymbol{w}}(\mathbf{z})
= \boldsymbol{\Phi}_{L-1}\!\left(
\boldsymbol{\Phi}_{L-2}\!\left(
\dots
\boldsymbol{\Phi}_0(\mathbf{z})
\right)\right)
\end{equation}

Each layer applies an affine transformation followed by a nonlinear activation. Because the training data only constrain the model at sampled points, the architecture affects how the NN behaves between those points and which solution it learns during training.

Concretely, each layer \(\boldsymbol{\Phi}_l\) is defined as
\(
\boldsymbol{\Phi}_l(\mathbf{z}_{l-1})
= \phi_l\!\bigl(\boldsymbol{W}_l\,\mathbf{z}_{l-1} + \mathbf{b}_l\bigr),
\)
where \(\boldsymbol{W}_l\) is the weight matrix, \(\mathbf{b}_l\) is the bias
vector, \(\mathbf{z}_{l-1}\) is the input to layer \(l\), and \(\phi_l\) is a
nonlinear basis function, usually called an activation function. By stacking such layers, \acp{NN}
construct hierarchical representations and efficiently approximate complex functions.
For compact notation, we collect all trainable weights and biases in
\(\boldsymbol{w}:=\{\boldsymbol{W}_l,\mathbf{b}_l\}_{l=0}^{L-1}\).

Ultimately, training a NN consists of determining, iteratively, the parameters
$\boldsymbol{w}$ that minimize a task-dependent objective or loss function \( \mathcal{L}(\mathrm{NN}_{\boldsymbol{w}}),
\) thereby selecting a function from the set of functions representable by the chosen
architecture. In the context of DAE systems, this objective must reflect the governing
physical constraints.

\subsection {Physics-Informed Machine Learning}
Building on the \ac{NN} representation above, we now formulate a
physics-informed solution operator for dynamic power-system components.

The solution operator, also called the flow map, maps a consistent initial condition and prescribed disturbances or control inputs to the system state at a future time.
To approximate the solution operator
\((\mathbf{x}_0,\mathbf{u}(\cdot),t_0,t)\mapsto\mathbf{x}(t_0+t)\) without
explicit time stepping at inference, we use a \ac{NN}:
\begin{equation}
\mathrm{NN}_{\boldsymbol{w}}(\mathbf{x}_0,\mathbf{u}(\cdot),t_0,t)
\approx \mathbf{x}(t_0+t)
\end{equation}

Rather than relying on labeled trajectories, we train
$\mathrm{NN}_{\boldsymbol{w}}$ by enforcing the DAE residual and the initial
condition as soft constraints. Let
$\{t_i\}_{i=1}^{N_t}$ denote the set of $N_t$ temporal collocation points, and
let $\{\mathbf{x}_{0,j},\mathbf{u}_j(\cdot)\}_{j=1}^{N_{\mathrm{ic}}}$ denote
the set of $N_{\mathrm{ic}}$ training scenarios, where
$\mathbf{x}_{0,j}$ is the initial state and $\mathbf{u}_j(\cdot)$ is the
corresponding input trajectory of scenario $j$. The network parameters
$\boldsymbol{w}$ are obtained by minimizing
\begin{align}
\label{eq:loss-compact}
\mathcal{L}
&=
\rho_{\mathrm{dae}}
\frac{1}{N_tN_{\mathrm{ic}}}
\sum_{i=1}^{N_t}\sum_{j=1}^{N_{\mathrm{ic}}}
\left\|
\mathbf{M}\partial_t\mathrm{NN}_{i,j}
-
f_{i,j}
\right\|_2^2
\nonumber\\
&\quad+
\rho_{\mathrm{ic}}
\frac{1}{N_{\mathrm{ic}}}
\sum_{j=1}^{N_{\mathrm{ic}}}
\left\|
\mathrm{NN}_{0,j}
-
\mathbf{x}_{0,j}
\right\|_2^2
\end{align}
where
\begin{align*}
\mathrm{NN}_{i,j}
&:=
\mathrm{NN}_{\boldsymbol{w}}
\left(
\mathbf{x}_{0,j},
\mathbf{u}_j(\cdot),
0,
t_i
\right),\\
\mathrm{NN}_{0,j}
&:=
\mathrm{NN}_{\boldsymbol{w}}
\left(
\mathbf{x}_{0,j},
\mathbf{u}_j(\cdot),
0,
0
\right),\\
f_{i,j}
&:=
f\left(
\mathrm{NN}_{i,j},
\mathbf{u}_j(t_i),
t_i
\right),\\
\partial_t\mathrm{NN}_{i,j}
&:=
\left.
\frac{\partial}{\partial t}
\mathrm{NN}_{\boldsymbol{w}}
\left(
\mathbf{x}_{0,j},
\mathbf{u}_j(\cdot),
0,
t
\right)
\right|_{t=t_i}
\end{align*}
The coefficients
$\rho_{\mathrm{dae}}$ and $\rho_{\mathrm{ic}}$ weight the DAE-residual and
initial-condition losses, respectively, while $\|\cdot\|_2$ denotes the
Euclidean norm.

Since $\mathrm{NN}_{\boldsymbol{w}}(\mathbf{x}_0,\mathbf{u}(\cdot),t_0,t)$ depends on the continuous variables $\mathbf{x}_0$, $\mathbf{u}(\cdot)$, and $t$, the problem formulation becomes
\begin{equation}
\mathbf{M}
\frac{\partial \mathrm{NN}_{\boldsymbol{w}}(\mathbf{x}_0,\mathbf{u}(\cdot),t_0,t)}
{\partial t}
-
f\!\left(
\mathrm{NN}_{\boldsymbol{w}}(\mathbf{x}_0,\mathbf{u}(\cdot),t_0,t),
\mathbf{u}(t),
t
\right)
=0,
\end{equation}
which defines a partial differential equation over the full input domain.
Training therefore requires representative sampling of \((\mathbf{x}_0,\mathbf{u},t_0,t)\) so that the learned operator produces
consistent trajectories across different initial conditions and inputs. When trajectory-simulated samples are available, a supervised data loss can be added to anchor the solution and ease the physics-informed optimization. In this work, we restrict attention to operators that take a single instantaneous 
input value $\mathbf{u}(t)$ and set $t_0 = 0$ without loss of generality. Although the formulation is developed for general index-1 DAE systems, PIML methods have predominantly been 
applied to Ordinary Differential Equations (ODEs), since DAEs introduce algebraic constraints that tightly couple state variables and complicate both training and numerical stability.

\subsubsection*{Geometric Interpretation of Integration}
\label{sec:geometry}

\begin{figure}[t]
\centering
\begin{tikzpicture}
\begin{axis}[
    width = 8cm,
    height = 4cm,
    axis lines = none,
    xmin = -0.2, ymin = -0.2,
    z buffer=sort, xmax = 1,
    zmax = 1.35,
    axis background/.style={fill=none},
]
    \coordinate (O) at (axis cs:-0.1,-0.1,-1);
    
    \draw[-latex, thick] (O) -- (axis cs:0.8,-0.1,-1)
    node[anchor=west] {$x_0$};
    
    \draw[-latex, thick] (O) --(axis cs:-0.1,0.3,-1)node[anchor=south] {$t$} ;
    
    \draw[-latex, thick] (O) -- (axis cs:-0.1,-0.1,1.5)
    node[anchor=west] {$\Phi(t,x_0)$};
    
    \addplot3 [surf,
        colormap/violet,
        domain=0:1, samples=45,
        line width=0.3pt
    ] {
      sin(deg(8*pi*x))*exp(-20*(y-0.5)^2)
    + 1.5*exp(-40*(x-0.5)^2
        - 8*(y-0.25)^2
        - (x-0.5)*(y-0.25))
    };

\def\xzero{0.1}

\addplot3[
    blue,
    thick,
    mark = *,
    samples=9,
    domain=0:0.6
]
(
    {\xzero},     
    {y},          
    {
        sin(deg(8*pi*\xzero))*exp(-20*(y-0.5)^2)
        + 1.5*exp(-40*(\xzero-0.5)^2
            - 8*(y-0.25)^2
            - (\xzero-0.5)*(y-0.25))
    }
);
\draw[-latex, thick, blue] (axis cs: 0.25,-0.2,-0.25) -- node[pos = 0, anchor=north  west, blue] {\scriptsize Single Time-Domain Trajectory} (axis cs:0.1,-0.05,-0.05);   

\end{axis}

\end{tikzpicture}
    
\caption{Flowmap view of a continuous DAE. The surface represents the flowmap
manifold. The curve shows the system's continuous trajectory. Discrete solver
steps move along this curve, approximating the underlying flow map.}
\label{fig:manifold}
\end{figure}
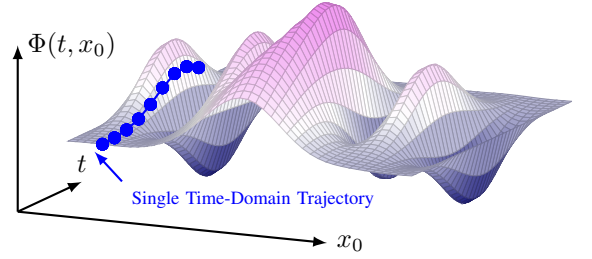

General DAE systems define a flow map whose trajectories lie on a
manifold. Time-stepping algorithms approximate the continuous evolution, starting from an initial condition, by moving along this manifold through a sequence of linear steps, according to the
local dynamics at the current point, as shown in Fig.~\ref{fig:manifold}.

In contrast, the PIML approaches aim to
approximate the flowmap itself as a global function, replacing step-by-step advancement with direct evaluation of the learned operator. Note that these PIML operators are often understood to be reduced order models. However, they do not project onto a lower-dimensional state. Instead, they can be understood as noisy replicas of the true system.

\subsection{Embedding Physical Knowledge into \ac{ML} Models}
Physics-informed losses enforce the governing equations during training, but they do not, on their own, determine how the \ac{ML} models interpolate between the training points to represent the underlying dynamics; that also depends on the architecture. In power-system applications, where oscillatory and multi-timescale behavior is often known a priori, it is therefore natural to embed such structure directly into the model design \cite{wang2020understandingmitigatinggradientpathologies}.

A representative example is the \ac{KAN} \cite{Ellinas2024,liu2024kankolmogorovarnoldnetworks, liu2025kan}, which replaces fixed activation functions such as \(\tanh\) with learnable univariate spline functions. This makes the model more adaptable during training and can help it capture the geometry of the flow-map manifold \cite{Ellinas2024}. However, spline evaluation and basis expansion make \acp{KAN} significantly more expensive at inference time. In power-system studies, where a surrogate may need to be evaluated repeatedly for many operating points or contingencies, this computational burden can become limiting.

The choice of activation function is central to this trade-off. Smooth monotone activations such as $\tanh$ train stably, whereas sinusoidal activations better match oscillatory behavior at the cost of more sensitive optimization. Although other periodic activations, e.g. \cite{liu2020snake}, may also be useful for power-system dynamics, their systematic evaluation is outside the scope of this work.

These considerations are directly relevant in power systems, where rotor-angle states are periodic and many transient responses contain oscillatory electromechanical or fast controller-driven modes. As a result, the activation choice affects not only training stability but also whether the surrogate reliably captures the system's dominant physical behavior.

These considerations motivate the use of architectures that retain the expressive
benefits of learnable activation functions while remaining computationally lightweight.

\subsubsection{ActNet Layers}
\begin{figure}[t]
    \centering
    \begin{tikzpicture}[node distance = 1.1cm]
        \node[circle, fill = black!10, draw = black](in){$\mathbf{x}_0$};
        \node[actnet1, right of = in,scale = 0.8, node distance = 2.5cm](n2){};
        \node[actnet2, above of = n2,scale = 0.8](n1){};
        \node[actnet3, below of = n2,scale = 0.8](n3){};

        \node[neuron, right of = n1, align = center, xshift = 0.5cm,scale = 0.8](n21){$s_1$};
        \node[neuron, right of = n2, align = center, xshift = 0.5cm,scale = 0.8](n22){$s_2$};
        \node[neuron, right of = n3, align = center, xshift = 0.5cm,scale = 0.8](n23){$s_3$};
        
        \node[right of = n22, circle, fill = black!10, draw = black, node distance = 2.5cm](out){$\hat{\mathbf{x}}_{NN}$};
    
        \draw[-latex] (in) --  (n1.west);
        \draw[-latex] (in) --  (n2.west);
        \draw[-latex] (in) --  (n3.west);

        \foreach \n in {n1, n2, n3}{
            \foreach \m in {n21,n22,n23}{
            \draw[-latex] (\n.east) --(\m.west);
            }
        }

        \path(n1.east) -- node[above, align = center,pos = 0.5]{\small $c_{ij}$} (n21.west);

        \draw[-latex](n21.east) --  (out);
        \draw[-latex](n22.east) -- (out);
        \draw[-latex](n23.east) --  (out);
        
        \begin{pgfonlayer}{bg}
            \path (in) -- +(+1.7,1.75) coordinate (r1);
            \path (out) -- +(-1.7,-1.75) coordinate (r2);
            \draw[pBlue, dashed,fill = pBlue!10] (r1) rectangle(r2);
    \end{pgfonlayer}
    \end{tikzpicture}
    
    
    \caption{ActNet layer structure. Each input variable is passed through trainable sinusoidal channels and mixed through coefficients $c_{ij}$, allowing the layer to represent timescales within a feedforward architecture.}
    \label{fig:actnet}
\end{figure}
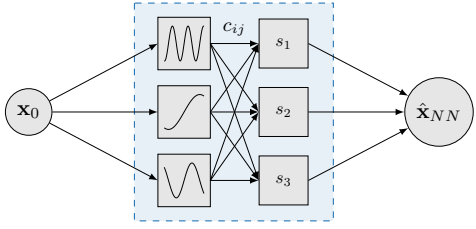

Figure~\ref{fig:actnet} illustrates the main idea of ActNet. Each input is
processed by several trainable sinusoidal channels, which can represent
different oscillation frequencies. The resulting channel responses are
weighted and combined to form the layer output, allowing slow and fast
oscillatory components to be represented simultaneously.

Besides \acp{KAN}, ActNet provides a lightweight alternative that embeds
oscillatory structure in its activation functions. It represents the
solution using trainable sinusoidal channels that can capture slow and
fast oscillatory components within the same architecture. Here, a channel denotes one trainable sinusoidal basis function. For each
channel \(i=1,\dots,m\), ActNet defines the univariate adaptive warp
\begin{equation}
    \phi_i(t) = \frac{\sin(\omega_i t + p_i) - \mu_i}{\sigma_i + \varepsilon},
\end{equation}
where \(\omega_i\) and \(p_i\) are trainable frequency and phase parameters, and
\(\mu_i,\sigma_i\) normalize the output for numerical stability. Since
\(\phi_i\) depends on trainable frequency and phase parameters, ActNet can be
interpreted as using learnable sinusoidal activation functions.

Given an input vector \(\mathbf{x}=(x_1,\ldots,x_n)\), ActNet forms the mixed
response of channel \(i\) as
\begin{equation}
    s_i(\mathbf{x}) = \sum_{j=1}^n c_{ij}\,\phi_i(x_j),
\end{equation}
and aggregates the channel responses as
\begin{equation}
    \mathrm{NN}_{\boldsymbol{w}}(\mathbf{x}) = \sum_{i=1}^m s_i(\mathbf{x}).
\end{equation}
Here, \(c_{ij}\) determines how strongly input variable \(x_j\) contributes to the
\(i\)-th sinusoidal channel. The parameter \(\omega_i\) determines the frequency of the \(i\)-th
sinusoidal channel, while \(c_{ij}\) weights the contribution of input
\(x_j\) to that channel. For ActNet,
\(
\boldsymbol{w}
:=
\{\omega_i,p_i\}_{i=1}^{m}
\cup
\{c_{ij}\}_{i=1,j=1}^{m,n}.
\)

This structure aligns the architecture with the multi-timescale nature of
power-system dynamics. Channels with smaller \(\omega_i\) can represent slowly
varying components, while channels with larger \(\omega_i\) can represent faster
oscillatory components. This is useful for component models in which electromechanical,
electrical, and fast-control interactions shape the dynamic response. Compared to spline-based KAN architectures, ActNet remains computationally lighter while
retaining a structure-aware representation for stiff and multi-timescale dynamics. The resulting layer structure is shown in Fig.~\ref{fig:actnet}.


\section{Neural Tangent Kernel and Training Dynamics} \label{sec:NTK}

The input--output mapping of a general \ac{NN} is highly nonlinear and
nonconvex in its parameters. Therefore, understanding how the trainable
parameters \(\boldsymbol{w}\) evolve and why training succeeds or fails
requires tools beyond classical approximation theory. In this work, we
adopt the NTK as a principled framework for analyzing how \acp{NN} adapt
their parameters during training.

The NTK for \acp{NN} is analogous to small-signal stability analysis of dynamic models. 
When a nonlinear dynamical system
\(\dot{\mathbf{x}} = F(\mathbf{x})\) is linearized around an operating point
\(\mathbf{x}^*\), the resulting state-space matrix
\(\mathbf{A} =
\left.\partial F/\partial \mathbf{x}\right|_{\mathbf{x}^*}\)
describes the local behavior of small perturbations around that operating point. Its eigenvalues determine whether modes decay,
oscillate, or grow, and at what rates. This modal interpretation provides the conceptual basis for comparing
physical-system modes with training-error modes.

NTK theory provides a similar modal view of \ac{NN} training. During training, gradient descent updates the trainable parameters \(\boldsymbol{w}\), which changes the NN output and reduces the training error. The NTK describes how sensitive the NN outputs at different training samples are to the parameter updates of $\boldsymbol{w}$, showing whether errors at different inputs or loss function terms reduce together or evolve independently during gradient descent.

Under the infinite-width NTK approximation, the \ac{NN} can be linearized around its initialization and the NTK is treated as approximately fixed during training. In this regime, the NTK plays the role of a training-domain state-space matrix: its eigenvectors define independent training-error, i.e, loss function, modes, and its eigenvalues determine their decay rates. This allows us to interpret \ac{NN} training using the same modal intuition used for power-system small-signal dynamics.

During training by
gradient descent on a loss \(\mathcal{L}(\mathrm{NN}_{\boldsymbol{w}})\), the parameters
follow the continuous gradient-flow dynamics
\begin{equation} \label{eq:gradient_flow}
\frac{d\boldsymbol{w}}{d\tau} = -\,\nabla_{\boldsymbol{w}}
\mathcal{L}(\mathrm{NN}_{\boldsymbol{w}}),
\end{equation}
where \(\nabla_{\boldsymbol{w}}\mathcal{L}\) denotes the gradient of the
loss with respect to the trainable parameter vector \(\boldsymbol{w}\), and
\(\tau\) denotes continuous training time, i.e., the continuous-time
analogue of training iterations or epochs.

Consider the mean-squared error loss
\[
\mathcal{L}(\mathrm{NN}_{\boldsymbol{w}})
= \tfrac12\sum_{i=1}^{N}
\bigl(\mathrm{NN}_{\boldsymbol{w}}(\mathbf{x}_i)-\mathbf{y}_i\bigr)^2,
\]
where \(\mathbf{y}_i\) denotes the target output at the training point \(\mathbf{x}_i\). The induced learning dynamics of the network output satisfy
\begin{equation}
\begin{aligned}
\frac{d}{d\tau}\mathrm{NN}_{\boldsymbol{w}}(\mathbf{x})
= -\sum_{i=1}^{N}
&\bigl(\mathrm{NN}_{\boldsymbol{w}}(\mathbf{x}_i)-\mathbf{y}_i\bigr) \\
&\times
\left\langle
\nabla_{\boldsymbol{w}} \mathrm{NN}_{\boldsymbol{w}}(\mathbf{x}_i),
\nabla_{\boldsymbol{w}} \mathrm{NN}_{\boldsymbol{w}}(\mathbf{x})
\right\rangle .
\end{aligned}
\end{equation}

The inner product
\begin{equation}
\bm{\Theta}(\mathbf{x}_i,\mathbf{x}_j)
= \left\langle\nabla_{\boldsymbol{w}} \mathrm{NN}_{\boldsymbol{w}}(\mathbf{x}_i),
         \nabla_{\boldsymbol{w}} \mathrm{NN}_{\boldsymbol{w}}(\mathbf{x}_j)\right\rangle
\end{equation}
is the NTK.

Collecting the predictions
on the training set into the vector
\(
\mathbf{U}(\tau)=
\bigl(\mathrm{NN}_{\boldsymbol{w}(\tau)}(\mathbf{x}_1),\dots,
\mathrm{NN}_{\boldsymbol{w}(\tau)}(\mathbf{x}_N)\bigr)^\top,
\)
the training error is \(\mathbf{e}(\tau)=\mathbf{U}(\tau)-\mathbf{y}\). Each eigenvector of the NTK describes a distinct way in which prediction
errors are distributed across the training samples, and the total training
error can be understood as a weighted sum of these error modes. The weights of this weighted sum
indicate how strongly each mode contributes to the current error, while the
eigenvalue \(a_i\) associated with each mode $i$ determines how quickly that contribution
is reduced during training: large eigenvalues indicate fast learning, whereas
small eigenvalues indicate slowly corrected or nearly unchanged errors. This
is analogous to expressing a physical disturbance as a combination of dynamic
modes in small-signal analysis.
In these modal coordinates, the training dynamics decouple into
independent scalar equations
\begin{equation}
    \frac{d a_i}{d\tau} = -\lambda_i\, a_i
\end{equation}
so each error component evolves independently as
\(a_i(\tau)=a_i(0)e^{-\lambda_i \tau}\).
Thus, the eigenvectors define independent \textit{error directions},
while the eigenvalues \(\lambda_i\) directly determine the decay rate
of each training-loss error mode: large eigenvalues lead to fast convergence, and small eigenvalues produce slow decay of the corresponding residual errors.

Under the infinite-width NTK approximation, the NTK is treated as approximately constant during training,
\(\bm{\Theta}(\mathbf{x}_i,\mathbf{x}_j)\approx
\bm{\Theta}_0(\mathbf{x}_i,\mathbf{x}_j)\)
\cite{jacot2020neuraltangentkernelconvergence,wang2020pinnsfailtrainneural}.
This means that, although the NN output changes, the sensitivity of the output to the trainable parameters changes only weakly. Under the fixed-NTK approximation, the output dynamics reduce to:
\begin{equation} \label{eq:ntk_constant}
\frac{d\mathbf{U}}{d\tau}
=
-\,\bm{\Theta}_0(\mathbf{U}-\mathbf{y}),
\end{equation}
so each training mode decays at a rate determined by the eigenvalues of the initial NTK. Please note that this is an approximation. In particular, when the parameters change substantially throughout training, the NTK values may also vary.

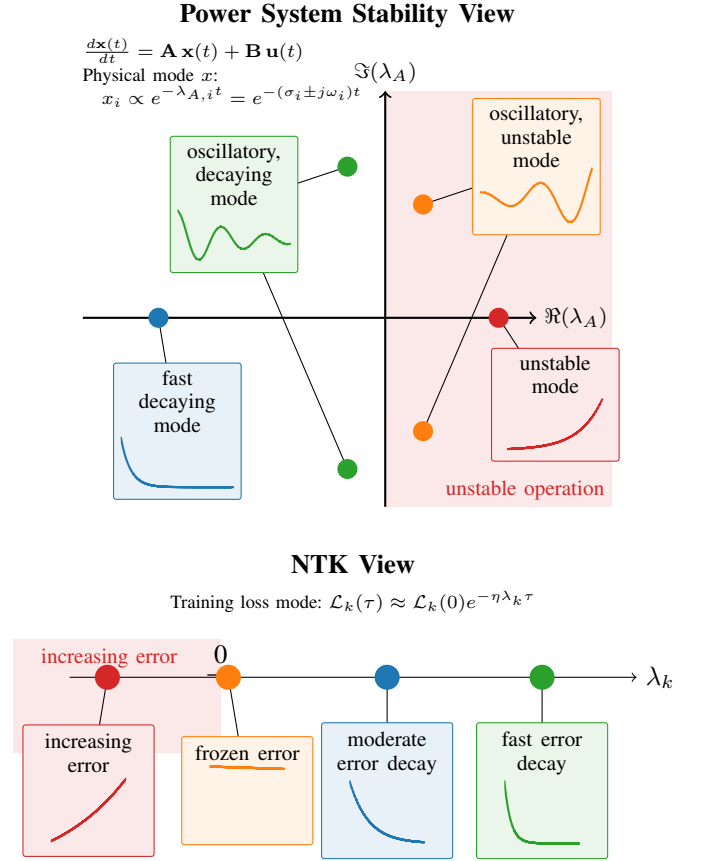
\begin{figure}[t]
  \centering
  \begin{tikzpicture}[node distance = 1.1cm]
\footnotesize
        \node at(-0.5,4){\normalsize \textbf{Power System Stability View}};
        \draw[draw = none, fill=sRed!10  ] (0,-2.5)rectangle(3,3);
        \node[anchor = south east, sRed] at(3,-2.5) {\footnotesize unstable operation};
        \draw[->, thick] (-4,0) -- (2,0) node[right]{$\Re(\lambda_A)$};
        \draw[->, thick] (0,-2.5) -- (0,3) node[above]{$\Im(\lambda_A)$};

        \node[text width = 6cm] at(-1,3.25){ \scriptsize
            \( \frac{d\mathbf{x}(t)}{dt} = \mathbf{A}\,\mathbf{x}(t)+\mathbf{B}\,\mathbf{u}(t) \) \\[0.15em]
            Physical mode $x$:\\
            \hspace{0.25cm}\(x_i \propto e^{-\lambda_{A,i} t} = e^{-(\sigma_i \pm j \omega_i)t} \)
       };
        
      \draw(-3,0) node[circle, fill = sBlue, sBlue, minimum size = 2pt]{} -- +(0.25,-1.5) node[block, fill = sBlue!10, draw = sBlue, text width = 1.5cm, align = center, rounded corners = true] {\footnotesize fast decaying mode\\ \centering
      \begin{tikzpicture}
        \begin{axis}[
            width=1.5cm,
            height=0.8cm,
            xmin=0, xmax=2,
            ymin=-0.1, ymax=1.1,
            axis lines=none,
            grid = none, 
            axis background/.style={fill=none},
        ]
        \addplot[sBlue, thick, domain=0:2, samples=50]
            {exp(-6*x)};
        \end{axis}
        \end{tikzpicture}};

       \draw (-0.5,2) node[circle, fill = sGreen, sGreen, minimum size = 2pt]{} -- +(-1.5,-0.5) node[block, fill = sGreen!10, draw = sGreen, text width = 1.5cm, align = center, rounded corners = true] (n1){\footnotesize oscillatory, decaying mode\\ \centering
      \begin{tikzpicture}
        \begin{axis}[
            width=1.5cm,
            height=0.8cm,
            xmin=0, xmax=2,
            axis lines=none,
            grid = none, 
            axis background/.style={fill=none},
        ]
        \addplot[sGreen, thick, domain=0:3, samples=50]
            {exp(-x)*2*cos(deg(8*x))};
        \end{axis}
        \end{tikzpicture}};
        \draw(-0.5,-2) node[circle, fill = sGreen, sGreen, minimum size = 2pt]{} -- (n1);

       \draw(1.5,0) node[circle, fill = sRed, sRed, minimum size = 2pt]{} -- +(0.75,-1.15) node[block, fill = sRed!10, draw = sRed, text width = 1.5cm, align = center, rounded corners = true] {\footnotesize unstable mode\\ \centering
      \begin{tikzpicture}
        \begin{axis}[
            width=1.5cm,
            height=0.8cm,
            axis lines=none,
            grid = none, 
            axis background/.style={fill=none},
        ]
        \addplot[sRed, thick, domain=0:2, samples=50]
            {exp(2*x)};
        \end{axis}
        \end{tikzpicture}};

      \draw (0.5,1.5) node[circle, fill = sOrange, sOrange, minimum size = 2pt]{} -- +(+1.5,+0.5) node[block, fill = sOrange!10, draw = sOrange, text width = 1.5cm, align = center, rounded corners = true](n2) {\footnotesize oscillatory, unstable mode\\ \centering
      \begin{tikzpicture}
        \begin{axis}[
            width=1.5cm,
            height=0.8cm,
            xmin=0, xmax=2,
            axis lines=none,
            grid = none, 
            axis background/.style={fill=none},
        ]
        \addplot[sOrange, thick, domain=0:3, samples=50]
            {exp(x)*2*cos(deg(6*x))};
        \end{axis}
        \end{tikzpicture}};
        \draw (0.5,-1.5)node[circle, fill = sOrange, sOrange, minimum size = 2pt]{} -- (n2); 
      \end{tikzpicture}
  
  
  \vspace{0.5cm}
  
  \begin{tikzpicture}[node distance = 1.1cm]
    \node at(1.75,1.5){\textbf{NTK View}};
    \draw[draw = none, fill=sRed!10  ] (-2.75,-1)rectangle(0,0.5);
    \node[anchor = north west, sRed] at(-2.5,0.5) {\footnotesize increasing error};
    \draw[->] (-2,0) --  (5.5,0) node[right]{$\lambda_k$};
    \node[]at(0,0) {|};
    \node[]at(0,0.3) {0};

   \node[] at(1.75,1){ \scriptsize
            Training loss mode: \( \mathcal{L}_k(\tau)\approx\mathcal{L}_k(0) e^{-\eta \lambda_k \tau }\)
       };

     \draw(4.25,0) node[circle, fill = sGreen, sGreen, minimum size = 2pt]{} -- +(0,-1.5) node[block, fill = sGreen!10, draw = sGreen, text width = 1.5cm, align = center, rounded corners = true] {\footnotesize fast error decay\\ \centering
      \begin{tikzpicture}
        \begin{axis}[
            width = 1cm,
            height=1cm,
            xmin=0, xmax=2,
            ymin=-0.1, ymax=1.1,
            axis lines=none,
            grid = none, 
            axis background/.style={fill=none},
        ]
        \addplot[sGreen, thick, domain=0:2, samples=50]
            {exp(-6*x)};
        \end{axis}
        \end{tikzpicture}};

        \draw(2.2,0) node[circle, fill = sBlue, sBlue, minimum size = 2pt]{} -- +(0,-1.5) node[block, fill = sBlue!10, draw = sBlue, text width = 1.5cm, align = center, rounded corners = true] {\footnotesize moderate error decay\\ \centering
      \begin{tikzpicture}
        \begin{axis}[
            width = 1cm,
            height=1cm,
            xmin=0, xmax=2,
            ymin=-0.1, ymax=1.1,
            axis lines=none,
            grid = none, 
            axis background/.style={fill=none},
        ]
        \addplot[sBlue, thick, domain=0:2, samples=50]
            {exp(-2*x)};
        \end{axis}
        \end{tikzpicture}};

        \draw(0.1,0) node[circle, fill = sOrange, sOrange, minimum size = 2pt]{} -- +(0.25,-1.5) node[block, fill = sOrange!10, draw = sOrange, text width = 1.5cm, align = center, rounded corners = true] {\footnotesize frozen error \\ \centering
      \begin{tikzpicture}
        \begin{axis}[
            width = 1cm,
            height=1cm,
            xmin=0, xmax=2,
            ymin=-0.1, ymax=1.1,
            axis lines=none,
            grid = none, 
            axis background/.style={fill=none},
        ]
        \addplot[sOrange, thick, domain=0:2, samples=50]
            {exp(-0.02*x)};
        \end{axis}
        \end{tikzpicture}};

        \draw(-1.5,0) node[circle, fill = sRed, sRed, minimum size = 2pt]{} -- +(-0.25,-1.5) node[block, fill = sRed!10, draw = sRed, text width = 1.5cm, align = center, rounded corners = true] {\footnotesize increasing error \\ \centering
      \begin{tikzpicture}
        \begin{axis}[
            width = 1cm,
            height=1cm,
            xmin=0, xmax=2,
            axis lines=none,
            grid = none, 
            axis background/.style={fill=none},
        ]
        \addplot[sRed, thick, domain=0:2, samples=50]
            {exp(0.5*x)};
        \end{axis}
        \end{tikzpicture}};
        
    \end{tikzpicture}
  
  \caption{Analogy between small-signal stability analysis and NTK training dynamics. Top: eigenvalues of the state-space matrix determine the behavior of physical perturbation modes. Bottom: NTK eigenvalues determine the decay rates of training-error modes. Large NTK eigenvalues correspond to fast error decay, while small eigenvalues correspond to slow or nearly frozen modes.}
  \label{fig:explain_eig}
\end{figure} 

Fig.~\ref{fig:explain_eig} summarizes the resulting parallel between power-system dynamic modes and training-error modes. In small-signal analysis, the eigenvalues of the linearized state-space matrix determine how dynamic modes behave after a system perturbation. The location of an eigenmode determines whether an oscillation occurs, at what rate, and whether it decays or grows. Similarly, the eigenvalues of the NTK determine how different training-error directions over the sampled points decay during the training process. An uneven NTK spectrum indicates optimization stiffness: some loss function error modes are corrected rapidly, while others remain nearly frozen. These slow NTK modes can limit the accuracy of the surrogate, especially for fast or oscillatory parts of the physical response. In contrast, an even NTK spectrum indicates that all errors decay at a similar rate and are learned with comparable speed and accuracy. Consequently, an ideal NTK profile would mean that NTK eigenmodes are similar across error components.

\subsection{NTK for PIML Models}
\label{subsec:PIML_NTK}

PIML training involves multiple loss terms that enforce data, initial
conditions, and physics constraints.

The coupling between the initial-conditions loss $\mathcal{L}_b(\mathcal{L}(\mathrm{NN}_{\boldsymbol{w}(\tau)}))$ and physics loss
$\mathcal{L}_p(\mathrm{NN}_{\boldsymbol{w}(\tau)}))$ from \eqref{eq:loss-compact} can be analyzed through the block NTK matrix
$\mathbf{K}$, so that their interaction is approximated by
\begin{equation}
\begin{bmatrix}
\dot{\mathcal{L}}_b(\mathrm{NN}_{\boldsymbol{w}(\tau)}))\\
\dot{\mathcal{L}}_p(\mathrm{NN}_{\boldsymbol{w}(\tau)}))
\end{bmatrix}
= 
-
\underbrace{
\begin{bmatrix}
\mathbf{K}_{bb} & \mathbf{K}_{bp} \\
\mathbf{K}_{pb} & \mathbf{K}_{pp}
\end{bmatrix}
}_{\mathbf{K}}
\begin{bmatrix}
\mathcal{L}_b(\mathrm{NN}_{\boldsymbol{w}(\tau)})) \\
\mathcal{L}_p(\mathrm{NN}_{\boldsymbol{w}(\tau)}))
\end{bmatrix},
\end{equation}
where each block is a Gram matrix of inner products between parameter gradients of the
corresponding loss terms. For example, if $i$ indexes samples in the initial conditions loss
and $j$ indexes samples in the physics loss:
\[
(\mathbf{K}_{bp})_{ij}
=
\left\langle
\frac{\partial \mathcal{L}_b(\mathrm{NN}_{\boldsymbol{w}(\tau)}(x_i^b))}{\partial\boldsymbol{w}},
\frac{\partial \mathcal{L}_p(\mathrm{NN}_{\boldsymbol{w}(\tau)}(x_j^p))}{\partial\boldsymbol{w}}
\right\rangle. 
\]
Here, $\mathbf{K}$ extends the NTK interpretation to the multi-objective setting.
Its diagonal blocks quantify how strongly each loss function term drives $\boldsymbol{w}$ updates, while the off-diagonal blocks quantify the interaction between different loss terms.
For each diagonal block, large leading eigenvalues correspond to training-error directions that respond strongly to changes in $\boldsymbol{w}$ and therefore decay rapidly under gradient descent. A long tail of small eigenvalues indicates error directions that respond weakly to parameter updates, so they decay slowly and may limit convergence. Since the trace of the matrix, $\mathrm{Tr}(K_{ii})$ equals the sum of the eigenvalues of the $i$-th block, it measures the overall strength with which the loss term $i$ affects the training update.
Large differences in $\mathrm{Tr}(K_{ii})$ therefore indicate imbalanced training, where some objectives dominate while others evolve slowly.

\subsection{NTK Eigenvalues for Adaptive Weights During Training}
Since the NTK eigenmodes provide insights into how fast each loss component is learned during training, they can be used to update the training weights during gradient descent. Guided by the NTK analysis, we formulate an adaptive weighting scheme that promotes uniform convergence across the individual loss objectives. Each loss term is assigned a weight \(\rho_i\), chosen inversely to the trace of its corresponding self-block:
\begin{equation} \label{eq:adaptive_loss}
\rho_i = \frac{\sum_j \mathrm{Tr}(K_{jj})}{\mathrm{Tr}(K_{ii})},
\end{equation}
where \(K_{ii}\) is the diagonal block associated with the \(i\)-th loss. A large trace indicates that the loss strongly drives the parameter update, while a small trace indicates a weak contribution. The inverse-trace weighting acts as a training preconditioner: dominant objectives are down-weighted and weaker objectives are up-weighted, promoting more balanced convergence across loss terms.

\section{Case Studies} \label{sec:casestudies_results}
In this section, we evaluate vanilla PINNs, KANs, and ActNet on power system time-domain simulation tasks with increasing stiffness and timescale separation. We first report standard accuracy and runtime metrics, and then use the NTK analysis to interpret the results in terms of training-error modes and loss-term imbalance. The goal is to connect physical stiffness in the underlying power-system dynamic model with optimization stiffness in PIML training, and to assess how different architectures represent oscillatory, stiff, and multi-timescale dynamics. We analyze three sets of power system test cases to illustrate three qualitatively different regimes:

\begin{enumerate}
\item \textbf{Mildly Stiff Dynamics (Synchronous Generator Models):} 
We first use classic \ac{SM} models as baseline test cases to compare vanilla PINNs, KANs, and ActNet in terms of accuracy and runtime in a regime where stiffness is limited.
\item \textbf{Stiff System with Moderate Time-Scale Separation (Modified Single Machine Infinite Bus (SMIB)):} Next, we modify a classic SMIB system to introduce a fast electrical mode and examine whether the architectures can simultaneously learn slow and fast dynamics. The performed NTK analysis provides insights into the performance of the \acp{NN}.
\item \textbf{Strongly Multiscale and Stiff Dynamics (Inverter Model):} Finally, we attempt to learn an 11th-order inverter model with nested control loops that induce both stiffness and pronounced timescale separation. Again, we demonstrate how the NTK analysis offers valuable insights into the model performance. Note that to the best of our knowledge, there are no reports in the literature of accurately learning such a stiff DAE system with \acp{NN}.
\end{enumerate}

The test system, implementation details, training configuration, and resulting performance are analyzed per scenario below, emphasizing how stiffness and multi-timescale structure influence learnability across architectures. Code and supplementary material are available at \cite{githubGitHubElpetros99tools_explaining_power}.

\subsection{Regime 1: Classic Synchronous-Machine Models}
\subsubsection{Test Case}
We first compare the accuracy and runtime of the three architectures on the 2nd (2D), 4th (4D), and 6th-order (6D) \ac{SM} models formulated in \cite{githubGitHubElpetros99tools_explaining_power}. These models provide a controlled setting in which the dynamical complexity increases with model order. To keep the training problem tractable, the experiments were carried out on a restricted set of initial conditions around the nominal operating region, including in the training 200 simulated trajectories, including an extra term in the loss function \cite{Ellinas2024}.

\subsubsection{Implementation and Training}
We compare three physics-informed architectures: (i) a vanilla PINN, (ii) a KAN, and (iii) ActNet. For each \ac{SM} order (2D, 4D, and 6D), one network is trained for each architecture, resulting in nine trained models in total. The same architecture settings are used across all \ac{SM} implementations per \ac{NN} type to isolate the effect of model order and architecture type. Specifically, the vanilla PINN uses 3 hidden layers with 64 neurons per layer. ActNet uses the same depth and width as the vanilla PINN, with $m=4$ trainable sinusoidal channels. The KAN uses a single hidden layer with 20 neurons, providing a compact spline-based alternative. All models are trained with the same physics-informed objective for a fair comparison.

Training uses the SOAP optimizer \cite{vyas2025soapimprovingstabilizingshampoo} with learning rate $10^{-4}$.  As numerical baselines, we report runtimes from a CPU-based Runge-Kutta 5 (RK5) solver implemented in Julia and from a GPU-based solver implemented through \texttt{DiffEqGPU.jl}. All GPU-runtime measurements are reported on the same NVIDIA T4 GPU. Each runtime corresponds to a full trajectory calculation over a horizon of \(T=1\,\mathrm{s}\) for 1000 scenarios.

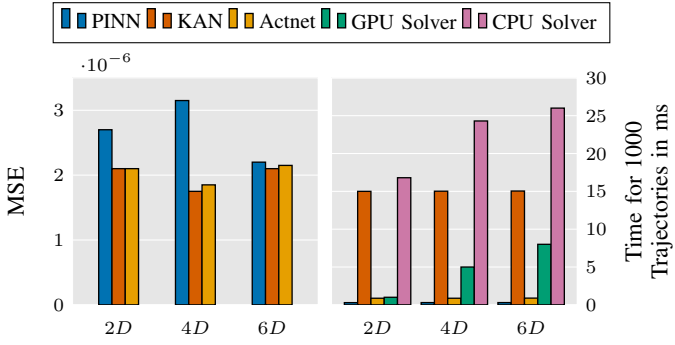
\begin{figure}[t]
  \centering
   \begin{tikzpicture}
\small
\begin{groupplot}[
    group style={
        group name=mygroup, 
        group size=2 by 3,
        vertical sep=0.8cm,
        horizontal sep=1.0cm,
        xticklabels at=edge bottom,
        xlabels at=edge bottom,
        vertical sep=15pt,
        horizontal sep=5pt,
    },
    xlabel={},
    width=3.25cm, 
    height=3cm,
    every node near coord/.style={
        /pgf/number format/1000 sep=,
        black,
        font=\tiny,
        /pgf/number format/precision=1,
        anchor=west,
        xshift=-10pt,
        yshift=5pt
    },
    nodes near coords align={left},
    ylabel style={text width=2.5cm, align=center},
    xmajorgrids=false,
    ylabel near ticks,
    enlarge x limits=0.3,
    scaled y ticks=true,
    ylabsh=-1.5cm,
    ylabel style={text width=3cm, align=center},
    every axis plot/.append style={line width=0.5pt}
]

\nextgroupplot[  
    ybar=0pt,
    ylabel={MSE},
    symbolic x coords={$2D$,$4D$,$6D$},
    xtick=data,
    ymin=0,
    ymax=3.5e-6,
    bar width=5pt,
    ytick distance=1e-6,
    ylabsh=-0.75cm
]
\addplot+[black, fill=cbBlue]       coordinates {($2D$,0.0000027)  ($4D$,0.00000315) ($6D$,0.0000022)};
\addplot+[black, fill=cbVermillion] coordinates {($2D$,0.0000021)  ($4D$,0.00000175) ($6D$,0.0000021)};
\addplot+[black, fill=cbOrange]     coordinates {($2D$,0.0000021)  ($4D$,0.00000185) ($6D$,0.00000215)};

\nextgroupplot[  
    ybar=0pt,
    ylabel={Time for 1000\\ Trajectories in ms},
    symbolic x coords={$2D$,$4D$,$6D$},
    xtick=data,
    ymin=0,
    ymax=30,
    bar width=5pt,
    ylabsh=4.2cm,    
    yticklabel pos=right,
    ylabel style={text width=3cm, align=center},
        legend style={
        at={(0,1.15)}, 
        anchor=south,
        legend columns=5
    },
]
\addplot+[black, fill=cbBlue]       coordinates {($2D$,0.29305)  ($4D$,0.3)      ($6D$,0.3)};
\addplot+[black, fill=cbVermillion] coordinates {($2D$,15.0)     ($4D$,15.02)    ($6D$,15.043)};
\addplot+[black, fill=cbOrange]     coordinates {($2D$,0.873537) ($4D$,0.873537) ($6D$,0.89)};
\addplot+[black, fill=cbGreen]      coordinates {($2D$,1.0)      ($4D$,5.0)      ($6D$,8.0)};
\addplot+[black, fill=cbPurple]     coordinates {($2D$,16.8)     ($4D$,24.3)     ($6D$,26.0)};
\legend{PINN, KAN, Actnet, GPU Solver, CPU Solver}

\end{groupplot}
\end{tikzpicture}
  \caption{Accuracy and runtime comparison for the different architectures on the three \ac{SM} test cases. Runtime is compared for evaluating 1000 trajectories and benchmarked against classic DAE GPU and CPU solvers.}
  \label{fig:plot_2d}
\end{figure}

\subsubsection{Results}

Figure~\ref{fig:plot_2d} summarizes the MSE and runtime of the three models on the 2D, 4D, and 6D \ac{SM} test cases, while also detailing the runtimes of both a GPU-based numerical solver and a CPU-based numerical solver. KAN and ActNet reduce the trajectory error relative to the vanilla PINN in the 2D and 4D test cases, with the largest reduction observed in the 4D case. At the same time, all three architectures achieve errors of the same order in the 6D case. Note that ActNet achieves these gains with a parameter count comparable to the vanilla PINN, suggesting that the improvement comes from the design of the activation functions rather than from a simple increase in model size.

The main difference between KAN and ActNet appears in inference cost. On the NVIDIA T4 GPU, the vanilla PINN is the fastest model, followed by ActNet, whereas KAN is substantially slower. Across all three test cases, ActNet is about $17\times$ faster than KAN. 

Comparing runtimes, ActNet is slightly faster than the GPU-based solver in the 2D case and up to about $9\times$ faster in the 6D case, while being nearly $30\times$ faster than the CPU-based solver benchmark. KAN, on the other hand, is slower than the GPU solver. This slowdown of KAN is structural: spline evaluation and basis expansion reduce the degree of vectorization available to KAN, while ActNet remains a fully feedforward architecture.

Overall, this baseline comparison shows that architectures with adaptive activation structures, namely KAN and ActNet, improve accuracy over the vanilla PINN. Among them, ActNet provides the best accuracy-runtime trade-off as it offers a computationally lightweight feedforward structure while using learnable sinusoidal activations.

\subsection{Regime 2: Capturing Slow and Fast Dynamics}
We next introduce stiffness into a classic 2nd-order SMIB model to examine if a single \ac{NN} architecture can capture both slow and fast modes. We therefore augment the \(2D\) SMIB model with terminal-voltage dynamics, obtaining a \(3D\) system with an additional fast electrical mode and the resulting stiffness. The GPU solver evaluates in approximately \(12\,\mathrm{ms}\) for 1000 scenarios, making it a promising candidate for PIML-based speedup.

\subsubsection{The Modified SMIB System}
To introduce one fast electrical mode into the 2nd-order SMIB dynamics, the terminal  voltage~$V(t)$ is introduced as a dynamic state, resulting in the following 3rd-order ODE system: 
\begin{align}
\begin{bmatrix}
1 & 0 & 0 \\
0 & H & 0 \\
0 & 0 & \tau
\end{bmatrix}
\frac{d}{dt}
\begin{bmatrix}
\delta \\
\omega \\
V
\end{bmatrix}
=
\begin{bmatrix}
\omega \\
P_m - E V \sin(\delta)+P_L(t)- D \omega \\
V_{\mathrm{ref}} - V + k \sin(\delta)
\end{bmatrix} \notag
\end{align}
Here, \(\delta\) is the rotor angle relative to the synchronous reference,
\(\omega\) is the speed deviation, and \(V\) is the terminal-voltage magnitude;
\(k\) controls the voltage-oscillation amplitude, while \(\delta\) sets its phase. Note that all symbol definitions and parameter values are provided in the supplementary material.

 A small time constant \(\tau\) introduces a fast electrical mode that creates stiffness in the ODE system and introduces a clear timescale separation: Linearizing around $(\delta,\omega,V)=(0,0,1)$ highlights that the voltage state $V$
contributes the fast electrical eigenvalue, while the rotor states
$\delta$ and $\omega$ yield the slow electromechanical eigenvalue $-D/(2H)$.
From these modal time constants, the corresponding stiffness ratio is:
$r \approx 2H/(D\tau) \approx 2\times 10^{3}$, indicating a difference of three orders of
magnitude between the fast and slow dynamics. Importantly, this stiffness is essentially one-dimensional, arising from a single dominant fast electrical mode.

\subsubsection{Implementation and Training}
We train three models: a vanilla PINN with $\tanh$ activations, a KAN, and an ActNet, from a single initial condition to isolate architectural and training effects from generalization across initial conditions. All models use the physics-informed loss in~\eqref{eq:loss-compact} and are trained for \num{5000} epochs using Adam~\cite{kingma2015adam}. We observe that the proposed adaptive weighting in~\eqref{eq:adaptive_loss} accelerates convergence and yields a lower final loss than unweighted training.  The vanilla PINN has two hidden layers of 32 neurons, the KAN has one hidden layer of 9 splines with grid size $G=7$ and spline order $k=3$, and the ActNet has two hidden layers of 5 neurons with $m=2$ trainable sinusoidal channels. These architectures were selected through a small grid search and then fixed for a fair comparison.

\begin{table}[t]
  \centering
  \scriptsize
  \renewcommand{\arraystretch}{1.2}
    \caption{Final loss values for ActNet, KAN, and the vanilla PINN on the modified SMIB test case.}
  \label{tab:loss-comparison}
    \begin{tabular}{lccc}
      \toprule
      \textbf{Loss Term} & \textbf{ActNet} & \textbf{KAN} & \textbf{PINN} \\
      \midrule
      $L_{\delta}$           & \cellcolor{sGreen!25} $4.27\times10^{-5}$    & \cellcolor{sGreen!25} $9.91\times10^{-5}$    & $3.93\times10^{-2}$    \\
      $L_{\omega}$           & \cellcolor{sGreen!25} $1.01\times10^{-5}$    & $2.95\times10^{-2}$    & \cellcolor{sRed!25}$2.25\times10^{-1}$    \\
      $L_{V}$           & $3.89\times10^{-3}$    & $4.44\times10^{-2}$    & \cellcolor{sRed!25} $ 4.05\times10^{-1}$    \\
      $L_{\mathrm{ic}}$ & \cellcolor{sGreen!25} $8.54\times10^{-8}$    & \cellcolor{sGreen!25}$3.24\times10^{-5}$    & $4.49\times10^{-2}$    \\
      \midrule
      $\displaystyle L_{\mathrm{tot}}$ 
                        & $3.94\times10^{-3}$    & $7.41\times10^{-2}$    & \cellcolor{sRed!25}$7.14\times10^{-1}$    \\
      \bottomrule
    \end{tabular}

\end{table}

\begin{figure}[t]
\centering
\begin{tikzpicture}
\begin{groupplot}[
    group style={
    group name = mygroup, 
    group size=1 by 1,          
    xticklabels at = edge bottom,
    vertical sep = 5pt
    },
    ylabsh=-2.7em,                   
    xlabel={},    
    xmin = 0, xmax = 4,    
    xtick distance = 0.5,         
    legend columns = 4,
    height=2.5cm,
    legend style={at={(1,1.17)}, anchor = east}
]

\nextgroupplot[                     
    ylabel={$\omega$ in rad/s},         
    xlabel = {time in s},
    scaled y ticks=false,
    yticklabel style={
        /pgf/number format/fixed,
        /pgf/number format/precision=5
    }
    ]
    \plot[black, thick] table[x=t, y=omega]{data/high_frequency/ground_predictions_eigenvalues.dat};     
    \addlegendentry{Radeau Solver}
    
    \plot[sBlue, thick] table[x= t, y=omega]{data/high_frequency/pinn_predictions_eigenvalues.dat};   
    \addlegendentry{PINN}
     
    \plot[sRed, dashed,thick] table[x= t, y=omega]{data/high_frequency/kan_predictions_eigenvalues.dat};     
    \addlegendentry{PI-KAN}

    \plot[sGreen,dashed, thick] table[x=t, y=omega]{data/high_frequency/actnet_predictions_eigenvalues.dat};     
    \addlegendentry{ActNet}

\end{groupplot}
\end{tikzpicture}
\caption{Predicted trajectories for the modified SMIB test case, comparing the vanilla PINN, KAN, and ActNet.}
\label{fig:plot_high_freq}
\end{figure}
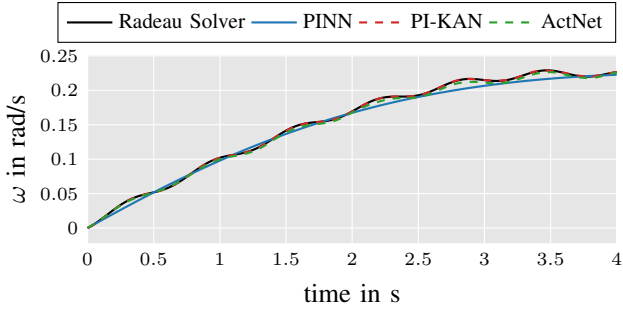 

    
    

\begin{figure}[h!]
\begin{minipage}{\columnwidth}
    \centering

    \begin{tikzpicture}
  \begin{groupplot}[
      group style = {
        group name = mygroup,
        group size = 1 by 3,          
        vertical sep = 5pt        
      },
      ylabsh = -3.25em,               
      xmin = 0, xmax = 100,          
      ymode = log,                   
      legend style = {
        at = {(0.5,1.05)},
        anchor = south,
        /tikz/every even column/.append style={column sep=0.25cm}
      },
      legend columns = 5,
      height = 1.5cm,
      width = 7cm, 
      ymin = 1e-17,
      ymax= 1e10
  ]

    \nextgroupplot[
      ylabel = {},
      xticklabels ={}
    ]
            
        \addlegendimage{empty legend}
        \addlegendentry{\textbf{Loss Term}:}
      \addplot[black, thick, only marks, mark = +, mark size = 1pt] table[x=ind, y expr=abs(\thisrow{K_11})] {data/high_frequency/eigenvalues_pinn2.dat};
      \addlegendentry{$L_\delta$}
      \addplot[sRed, thick, only marks, mark = +, mark size = 1pt] table[x=ind, y expr=abs(\thisrow{K_22})] {data/high_frequency/eigenvalues_pinn2.dat};
      \addlegendentry{$L_\omega$}
      \addplot[sGreen, thick, only marks, mark = +, mark size = 1pt] table[x=ind, y expr=abs(\thisrow{K_33})] {data/high_frequency/eigenvalues_pinn2.dat};
      \addlegendentry{$L_V$}
      \addplot[sBlue, thick, only marks, mark = +, mark size = 1pt] table[x=ind, y expr=abs(\thisrow{K_44})] {data/high_frequency/eigenvalues_pinn2.dat};
      \addlegendentry{$L_{ic}$}
      \node [below left, fill = white, opacity = 0.5, text opacity=1]  at (axis cs:99,1e10) {\footnotesize PINN};

    \nextgroupplot[
      ylabel = {Eigenvalue Magnitude},
      xticklabels ={}
    ]
      \addplot[black, thick, only marks, mark = +, mark size = 1pt] table[x=ind, y expr=abs(\thisrow{K_11})] {data/high_frequency/eigenvalues_kan2.dat};
      \addplot[sRed, thick, only marks, mark = +, mark size = 1pt] table[x=ind, y expr=abs(\thisrow{K_22})] {data/high_frequency/eigenvalues_kan2.dat};
      \addplot[sGreen, thick, only marks, mark = +, mark size = 1pt] table[x=ind, y expr=abs(\thisrow{K_33})] {data/high_frequency/eigenvalues_kan2.dat};
        \addplot[sBlue, thick, only marks, mark = +, mark size = 1pt] table[x=ind, y expr=abs(\thisrow{K_44})] {data/high_frequency/eigenvalues_kan2.dat};

      \node [below left, fill = white, opacity = 0.5, text opacity=1]  at (axis cs:99,5e8) {\footnotesize KAN};

    \nextgroupplot[
      ylabel = {},
      xlabel={Eigenvalue rank},
      legend to name = {} 
    ]
      \addplot[black, thick, only marks, mark = +, mark size = 1pt] table[x=ind, y expr=abs(\thisrow{K_11})] {data/high_frequency/eigenvalues_actnet2.dat};
      \addplot[sRed, thick, only marks, mark = +, mark size = 1pt] table[x=ind, y expr=abs(\thisrow{K_22})] {data/high_frequency/eigenvalues_actnet2.dat};
      \addplot[sGreen, thick, only marks, mark = +, mark size = 1pt] table[x=ind, y expr=abs(\thisrow{K_33})] {data/high_frequency/eigenvalues_actnet2.dat};
      \addplot[sBlue, thick, only marks, mark = +, mark size = 1pt] table[x=ind, y expr=abs(\thisrow{K_44})] {data/high_frequency/eigenvalues_actnet2.dat};
            \node [below left, fill = white, opacity = 0.5, text opacity=1]  at (axis cs:99,1e9) {\footnotesize Actnet};

  \end{groupplot}
\end{tikzpicture}
    \caption{Unweighted NTK eigenvalue spectra of the
    \(L_{\delta}\), \(L_{\omega}\), \(L_V\), and \(L_{\mathrm{ic}}\)
    loss blocks for the three architectures. The horizontal axis gives the
    eigenvalue rank after sorting by decreasing magnitude.}
    \label{fig:eigenvalues}

    \vspace{10pt}

    \begin{tikzpicture}
  \begin{groupplot}[
      group style = {
        group name = mygroup,
        group size = 1 by 3,          
        vertical sep = 5pt        
      },
      ylabsh = -3.25em,               
      xmin = 0, xmax = 100,          
      ymode = log,                   
      legend style = {
        at = {(0.5,1.05)},
        anchor = south,
        /tikz/every even column/.append style={column sep=0.25cm}
      },
      legend columns = 5,
      height = 1.5cm,
      width = 7cm, 
      ymin = 1e-17,
      ymax= 1e10
  ]

    \nextgroupplot[
      ylabel = {},
      xlabel={Eigenvalue rank},
      xticklabels= {}
    ]
         \addlegendimage{empty legend}
     \addlegendentry{\textbf{Loss Term}:}
      \addplot[black, thick, only marks, mark = *, mark size = 1pt] table[x=ind, y expr=abs(\thisrow{K_11})] {data/high_frequency/eigenvalues_pinn2_scaled.dat};
      \addlegendentry{$L_\delta$}
      \addplot[sRed, thick, only marks, mark = +, mark size = 1pt] table[x=ind, y expr=abs(\thisrow{K_22})] {data/high_frequency/eigenvalues_pinn2_scaled.dat};
      \addlegendentry{$L_\omega$}
      \addplot[sGreen, thick, only marks, mark = +, mark size = 1pt] table[x=ind, y expr=abs(\thisrow{K_33})] {data/high_frequency/eigenvalues_pinn2_scaled.dat};
      \addlegendentry{$L_V$}
      \addplot[sBlue, thick, only marks, mark = +, mark size = 1pt] table[x=ind, y expr=abs(\thisrow{K_44})] {data/high_frequency/eigenvalues_pinn2_scaled.dat};
      \addlegendentry{$L_{ic}$}
    \node [below left, fill = white, opacity = 0.5, text opacity=1]  at (axis cs:99,5e6) {\footnotesize PINN};

    \nextgroupplot[
      ylabel = {Eigenvalue Magnitude},
      xticklabels= {}
    ]
      \addplot[black, thick, only marks, mark = *, mark size = 1pt] table[x=ind, y expr=abs(\thisrow{K_11})] {data/high_frequency/eigenvalues_kan2_scaled.dat};
      \addplot[sRed, thick, only marks, mark = +, mark size = 1pt] table[x=ind, y expr=abs(\thisrow{K_22})] {data/high_frequency/eigenvalues_kan2_scaled.dat};
      \addplot[sGreen, thick, only marks, mark = +, mark size = 1pt] table[x=ind, y expr=abs(\thisrow{K_33})] {data/high_frequency/eigenvalues_kan2_scaled.dat};
        \addplot[sBlue, thick, only marks, mark = +, mark size = 1pt] table[x=ind, y expr=abs(\thisrow{K_44})] {data/high_frequency/eigenvalues_kan2_scaled.dat};
      \node [below left, fill = white, opacity = 0.5, text opacity=1]  at (axis cs:99,5e6) {\footnotesize KAN};

    \nextgroupplot[
      ylabel = {},
      xlabel={Eigenvalue rank},
      legend to name = {} 
    ]
      \addplot[black, thick, only marks, mark = *, mark size = 1pt] table[x=ind, y expr=abs(\thisrow{K_11})] {data/high_frequency/eigenvalues_actnet2_scaled.dat};
      \addplot[sRed, thick, only marks, mark = +, mark size = 1pt] table[x=ind, y expr=abs(\thisrow{K_22})] {data/high_frequency/eigenvalues_actnet2_scaled.dat};
      \addplot[sGreen, thick, only marks, mark = +, mark size = 1pt] table[x=ind, y expr=abs(\thisrow{K_33})] {data/high_frequency/eigenvalues_actnet2_scaled.dat};
      \addplot[sBlue, thick, only marks, mark = +, mark size = 1pt] table[x=ind, y expr=abs(\thisrow{K_44})] {data/high_frequency/eigenvalues_actnet2_scaled.dat};
        \node [below left, fill = white, opacity = 0.5, text opacity=1]  at (axis cs:99,5e6) {\footnotesize Actnet};

  \end{groupplot}
\end{tikzpicture}
    \caption{NTK eigenvalue spectra after applying the adaptive weights
    \(\rho_i\). The horizontal axis gives the eigenvalue rank after sorting
    by decreasing magnitude.}
    \label{fig:eigenvalues_scaled}

    \vspace{10pt}

    \begin{tikzpicture}
  \begin{groupplot}[
      group style = {
        group name = mygroup,
        group size = 1 by 1,          
        horizontal sep = 10pt        
      },
      ylabsh = -3em,               
      xmin = 0, xmax = 100,          
      ymode = log,                   
      legend style = {
        at = {(0.5,1.05)},
        anchor = south,
        /tikz/every even column/.append style={column sep=0.25cm}
      },
      legend columns = 5,
      height = 2cm,
      width = 7cm,                    
  ]

    \nextgroupplot[
      ylabel = {Eigv. for $K_{L_\delta}$},
      xlabel={Eigenvalue rank},
    ]
     \addlegendimage{empty legend}
     \addlegendentry{\textbf{No. of Iterations}:}
      \addplot[black, thick, only marks, mark = *, mark size = 1pt] table[x=ind, y expr=abs(\thisrow{K11_n0})] {data/high_frequency/eigenvalues_snapshots_actnet2.dat};
      \addlegendentry{0}
      \addplot[sRed, thick, only marks, mark = x, mark size = 2pt] table[x=ind, y expr=abs(\thisrow{K11_n10})] {data/high_frequency/eigenvalues_snapshots_actnet2.dat};
      \addlegendentry{1000}
      \addplot[sGreen, thick, only marks, mark = +, mark size = 2pt] table[x=ind, y expr=abs(\thisrow{K11_n20})] {data/high_frequency/eigenvalues_snapshots_actnet2.dat};
      \addlegendentry{2000}
      \addplot[sBlue, thick, only marks, mark = +, mark size = 1pt] table[x=ind, y expr=abs(\thisrow{K11_nend})] {data/high_frequency/eigenvalues_snapshots_actnet2.dat};
      \addlegendentry{3000}




  \end{groupplot}
\end{tikzpicture}
    \caption{ActNet NTK eigenvalue spectra at selected training checkpoints.
    Each curve represents the complete spectrum at the indicated iteration,
    with the horizontal axis denoting the eigenvalue rank.}
    \label{fig:eigenvalues_iterations}
\end{minipage}
\end{figure}

\subsubsection{Results}

Figure~\ref{fig:plot_high_freq} showcases one time-domain trajectory of the \ac{SM} frequency for the classic time-domain solver and the three \ac{NN} architectures. While the PINN fails to capture the faster oscillation around the main trajectory, both ActNet and PI-KAN successfully reproduce it. Note that the KAN appears more accurate.  

These findings are supported by the final physics-loss components across all evaluation trajectories presented in Table~\ref{tab:loss-comparison}. Here, \(L_{\delta}\), \(L_{\omega}\), and \(L_{V}\) denote the mean-squared residuals of the \(\delta\)-, \(\omega\)-, and \(V\)-equations, respectively; \(L_{\mathrm{ic}}\) denotes the initial-condition residual, and \(L_{\mathrm{tot}}\) the total loss. ActNet depicts the smallest total loss, more than one order of magnitude below KAN and nearly three below the vanilla PINN, with the largest gain in the residual associated with the oscillatory speed response. 

We next use the NTK spectra to analyze these differences, which occur despite applying the NTK-weighted training scheme. Generally,  the NTK spectrum indicates how different training-error modes respond to parameter updates: large leading eigenvalues correspond to modes that are learned quickly, while small tail eigenvalues correspond to modes that converge slowly. We therefore look for two features: how rapidly the spectrum decays, and whether the dominant eigenvalues are balanced across the residual losses. A sharp decay suggests that only a few modes are learned efficiently. In contrast, a milder decay with larger tail eigenvalues indicates that a broader range of modes remains trainable, including modes associated with faster or more oscillatory response components.

Figure~\ref{fig:eigenvalues} depicts the NTK spectra computed on 100 time-collocation points for the residual terms \(L_{\delta}\), \(L_{\omega}\), and \(L_V\) at the start of the training. The nearly flat \(L_{\mathrm{ic}}\) spectrum indicates similar correction rates because the same initial condition is enforced for all collocation samples at \(t=0\), without temporal-derivative terms. The vanilla PINN displays a very heavy tail, with eigenvalues decaying from about \(10^{10}\) to values close to zero. In contrast, KAN and ActNet show a milder decay with larger tail eigenvalues. The dominant eigenvalues are also more balanced across residual terms. In the vanilla PINN, \(L_{\delta}\) dominates near \(10^{10}\), while \(L_{\omega}\) and \(L_V\) remain near \(10^{1}\) to \(10^{2}\). For KAN and ActNet, the largest eigenvalues cluster around \(10^{8}\) to \(10^{9}\), with \(L_V\) slightly largest, producing a more even distribution of gradient contributions. This more balanced distribution, together with the larger tail eigenvalues, is
consistent with the time-domain results: KAN and ActNet learn the fast oscillatory component more effectively, whereas the vanilla PINN fails to capture it accurately.

Figure~\ref{fig:eigenvalues_scaled} shows the NTK spectra, at epoch 0, after applying the adaptive weighting rule formulated in~\eqref{eq:adaptive_loss}, which scales each loss term according to the inverse trace of its NTK block. This reduces disparities in the effective contribution of the residual terms and promotes more similar convergence rates across the state-equation residuals. Figure~\ref{fig:eigenvalues_iterations} compares the complete ActNet NTK
spectra at initialization and after 1000, 3000, and 5000 training iterations.
The horizontal axis gives the eigenvalue rank in decreasing order, while the
legend identifies the training checkpoint. Their strong overlap suggests that
the NTK remains approximately fixed during training.

\subsection{Regime 3: Inverter Model Benchmark}

The final case study considers an 11-state PQ-controlled, grid-following
inverter model. The GPU solver evaluates 1000 scenarios in approximately
\(27\,\mathrm{ms}\), making this benchmark relevant for acceleration.
Although its stiffness ratio,
\(r_{\mathrm{inv}}\approx1.7\times10^{2}\), is smaller than that of the
modified SMIB,
\(r_{\mathrm{SMIB}}\approx2\times10^{3}\),
the inverter contains several interacting current-control, power-control,
PLL, and filtering modes with different timescales
\cite{githubGitHubElpetros99tools_explaining_power}. We therefore examine
whether this distributed multi-timescale structure creates optimization
difficulties that are not captured by the stiffness ratio alone.

\subsubsection{Inverter Model and Training}

The inverter model includes a PLL, outer PQ control, inner current control,
current dynamics, and measurement filters, as detailed in
\cite{githubGitHubElpetros99tools_explaining_power}. We train a vanilla PINN,
ActNet, and KAN from a single initial condition for \(10{,}000\) epochs with
a learning rate of \(10^{-4}\), using the physics-informed objective
in~\eqref{eq:loss-compact}. The PINN and ActNet use three hidden layers,
while the KAN uses two hidden layers of 20 neurons each. For the NTK
comparison in Fig.~\ref{fig:eigenvalues_multiple}, we additionally evaluate
PINNs with widths 64, 128, 256, and 512. The PLL and outer-control states
evolve substantially more slowly than the current dynamics, producing large
differences in residual and gradient scales. We denote the residual loss of
state \(z\) by \(L_z\) and use a prediction horizon of \(0.1\,\mathrm{s}\),
unless stated otherwise.

\subsubsection{NTK-Based Diagnosis}

We diagnose the training difficulty at three levels: the per-residual NTK
spectra identify slowly learned error modes, the residual gradients reveal
imbalance between loss terms, and a short-horizon ablation separates
optimization difficulty from the ability of the NN to fit the trajectory.

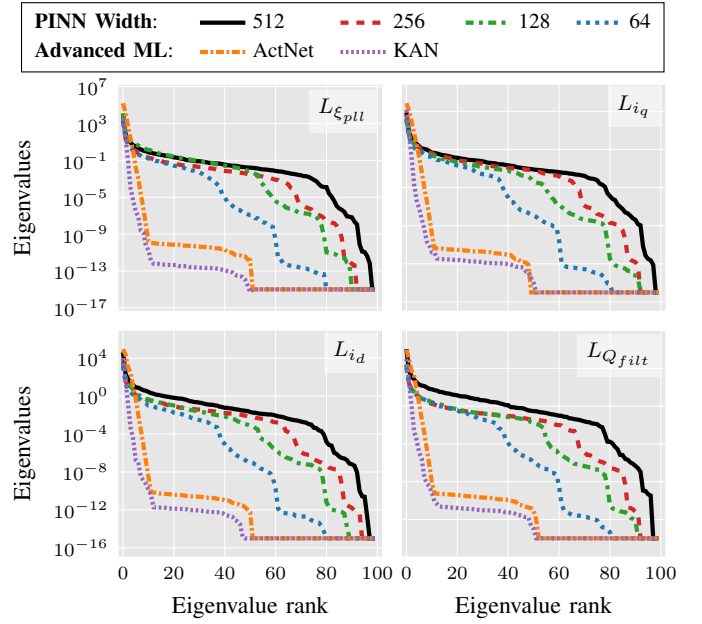
\begin{figure}[t]
    \centering
\begin{tikzpicture}
  \begin{groupplot}[
    group style={
      group name=ntkgroup,
      group size=2 by 2,      
      horizontal sep=0.25cm,
      vertical sep=0.25cm,    
      ylabels at=edge left,xlabels at=edge bottom, xticklabels at = edge bottom, yticklabels at = edge left
    },
    width=3.5cm, height=3cm,
    cycle list name=ntkstyles,
    xlabel={Eigenvalue rank},
    ylabel = {Eigenvalues},
    ymode=log,
    grid=both, grid style={dotted},
    tick label style={font=\small},
    label style={font=\small},
    title style={font=\footnotesize},
    legend style={
      at={(0.85,1.05)},
      anchor=south,
      draw=black,
      /tikz/every even column/.append style={column sep=0.25cm}
    },
    legend columns=5,
    xtick distance = 20,
  ]

    \pgfplotsinvokeforeach{1,2,3,9}{
      \nextgroupplot[
          xmin = -2, xmax = 102,
      ]
       \addlegendimage{empty legend}
        \addplot table[x expr=\coordindex, y index=#1] {images/inverter_ntk/eigvals_dim512.dat};
        \addplot table[x expr=\coordindex, y index=#1] {images/inverter_ntk/eigvals_dim256.dat};
        \addplot table[x expr=\coordindex, y index=#1] {images/inverter_ntk/eigvals_dim128.dat};
        \addplot table[x expr=\coordindex, y index=#1] {images/inverter_ntk/eigvals_dim64.dat};
        \addlegendimage{empty legend}
        \addplot table[x expr=\coordindex, y index=#1] {images/inverter_ntk/eigvals_actnet.dat};
        \addplot table[x expr=\coordindex, y index=#1] {images/inverter_ntk/eigvals_kang.dat};

        \ifnum#1=1
          \addlegendentry{\textbf{PINN Width}: }
          \addlegendentry{512}
          \addlegendentry{256}
          \addlegendentry{128}
          \addlegendentry{64}
        \addlegendentry{\textbf{Advanced ML}: }
          \addlegendentry{ActNet}
          \addlegendentry{KAN}
        \fi
        \node [below left, fill = white, opacity = 0.5, text opacity=1]  at (axis cs:100,5e6) {\footnotesize \ifcase#1\relax
            $L_{\theta_{pll}}$
          \or
            $L_{\xi_{pll}}$    
          \or
            $L_{i_q}$          
          \or
              $L_{i_d}$          
          \or
              $L_{\xi_{id}}$     
          \or
            $L_{\xi_{iq}}$       
          \or
          $L_{\xi_{P}}$
          \or
          $L_{\xi_{Q}}$
          \or
          $L_{P_{filt}}$
          \or
          $L_{Q_{filt}}$
          \or
            $L_{V_{filt}}$     
          \fi};
    }

  \end{groupplot}
\end{tikzpicture}
    \caption{NTK eigenvalue spectra of the diagonal loss blocks \(K_{i,i}\)
    for selected inverter residuals and different architectures. The
    residuals represent slow PLL dynamics, fast current dynamics, and
    intermediate filtering dynamics. The horizontal axis denotes the
    eigenvalue rank after sorting by decreasing magnitude.}
    \label{fig:eigenvalues_multiple}
\end{figure}

Figure~\ref{fig:eigenvalues_multiple} compares the slow PLL residual
\(L_{\xi_{\mathrm{pll}}}\), the fast current residuals \(L_{i_d}\) and
\(L_{i_q}\), and the intermediate filter residual
\(L_{Q_{\mathrm{filt}}}\). All architectures exhibit long tails of small
eigenvalues, meaning that many error modes are only weakly corrected by
parameter updates. Increasing the PINN width raises some eigenvalues but does
not remove these slowly learned modes, while ActNet and KAN show the same
qualitative limitation. Thus, increasing model size or changing among the
tested architectures does not by itself resolve the training difficulty.

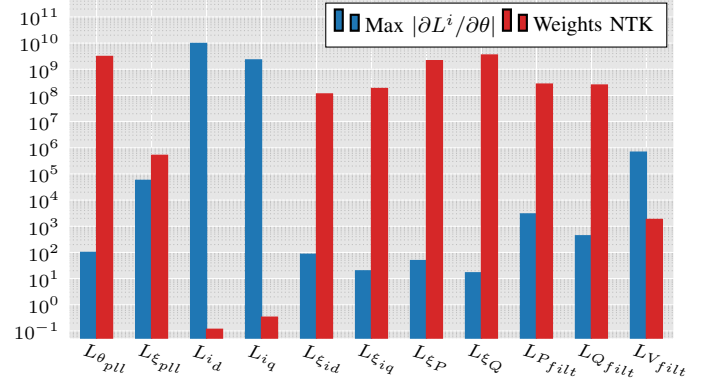
\begin{figure}[t]
    \centering
    \begin{tikzpicture}
  \begin{axis}[
    width=8cm,
    height=4.5cm,
    ymode=log,
    log origin=infty,
    ymin=5e-2, ymax=5e11,
    enlarge x limits=0.05,
    ybar,
    bar width=6.5pt,
    ylabel={},
    xlabel = {},
    symbolic x coords={
      {$L_{\theta_{pll}}$},{$L_{\xi_{pll}}$},{$L_{i_d}$},{$L_{i_q}$},
      {$L_{\xi_{id}}$},{$L_{\xi_{iq}}$},{$L_{\xi_{P}}$},{$L_{\xi_{Q}}$},
      {$L_{P_{filt}}$},{$L_{Q_{filt}}$},{$L_{V_{filt}}$}
    },
    xtick=data,
    x tick label style={rotate=-15,anchor=west, xshift = -8pt, yshift = -4pt},
    grid=both,
    grid style={densely dotted},
    legend style={
      at={(0.99,0.99)},
      anchor=north east,
      legend columns=2
    }
  ]
   \addlegendimage{no markers, fill = sBlue, draw = none, thick}
   \addlegendimage{no markers, fill = sRed, draw = none, thick}
    \addplot+[
      fill=sBlue,
      draw=none,
      bar shift=-3pt
    ] coordinates {
      ({$L_{\theta_{pll}}$},   1.061e+02)
      ({$L_{\xi_{pll}}$},       6.004e+04)
      ({$L_{i_d}$},             1.027e+10)
      ({$L_{i_q}$},             2.414e+09)
      ({$L_{\xi_{id}}$},        9.004e+01)
      ({$L_{\xi_{iq}}$},        2.095e+01)
      ({$L_{\xi_{P}}$},         5.160e+01)
      ({$L_{\xi_{Q}}$},        1.765e+01)
      ({$L_{P_{filt}}$},        3.139e+03)
      ({$L_{Q_{filt}}$},        4.587e+02)
      ({$L_{V_{filt}}$},        7.189e+05)
    };
    \addlegendentry{Max $|\partial L^i/\partial\theta|$}

    \addplot+[
      fill=sRed,
      draw=none,
      bar shift=3pt
    ] coordinates {
      ({$L_{\theta_{pll}}$},   3.27136631e+09)
      ({$L_{\xi_{pll}}$},       5.43436592e+05)
      ({$L_{i_d}$},             1.22263671e-01)
      ({$L_{i_q}$},             3.54555183e-01)
      ({$L_{\xi_{id}}$},        1.20777008e+08)
      ({$L_{\xi_{iq}}$},        1.94585054e+08)
      ({$L_{\xi_{P}}$},         2.25405312e+09)
      ({$L_{\xi_{Q}}$},         3.71210763e+09)
      ({$L_{P_{filt}}$},        2.87269003e+08)
      ({$L_{Q_{filt}}$},        2.64371826e+08)
      ({$L_{V_{filt}}$},        1.92943935e+03)
    };
    \addlegendentry{ Weights NTK}
    \legend{
      Max $|\partial L^i/\partial\theta|$,
      Weights NTK
    }
  \end{axis}
\end{tikzpicture}
    \caption{Maximum residual gradients with respect to the NN final-layer
    weights and the corresponding NTK-based weights from
    \eqref{eq:adaptive_loss}. The adaptive weights reduce the dominance of
    residuals with large gradients.}
    \label{fig:gradient_imbalance}
\end{figure}

Figure~\ref{fig:gradient_imbalance} examines the imbalance between residual
losses. The current residuals \(L_{i_d}\) and \(L_{i_q}\) produce the largest
gradients and therefore receive the smallest adaptive weights, while weaker
residuals receive larger weights. This weighting rebalances the influence of
the complete loss terms, but each weight scales all modes of a loss together
and therefore does not remove its small-eigenvalue tail.

\subsubsection{Short-Horizon Ablation}

To determine whether the remaining error arises from limited trajectory
representation or from the physics-informed optimization, we train a
simplified one-layer NN over \(10^{-4}\,\mathrm{s}\) using five
configurations: data loss only; physics loss with and without NTK weighting;
and combined data and physics losses with and without NTK weighting.

\begin{figure}[t]
    \centering
    \includegraphics[width=\linewidth]{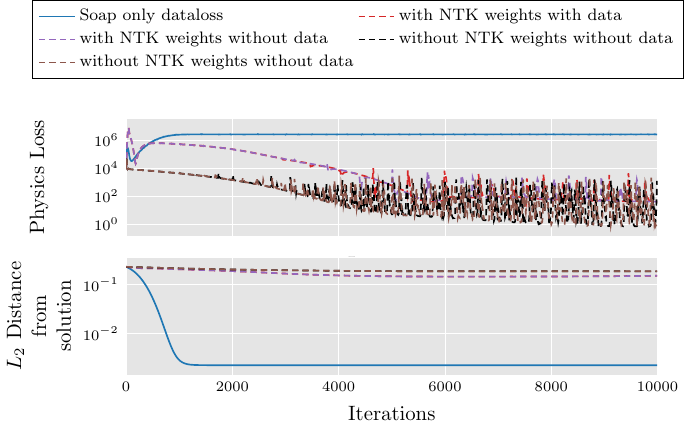}
    \caption{Short-horizon ablation comparing supervised and
    physics-informed training. The data-only model provides a supervised
    reference, while NTK weighting improves the combined data-and-physics
    objective relative to its unweighted counterpart.}
    \label{fig:inverter_loss}
\end{figure}

Figure~\ref{fig:inverter_loss} shows that the data-only model achieves a much
lower trajectory error, indicating that, over this short horizon, fitting
direct state observations is easier than satisfying the derivative-based
physics residuals. The NN can therefore represent the local trajectory,
while the physics-informed objective creates the main optimization
difficulty. NTK weighting improves the combined data-and-physics case
relative to unweighted training, although it remains less accurate than
direct supervised fitting. Nevertheless, physics-informed training provides
an explicit and physically interpretable objective and avoids reliance on
labeled trajectories generated by numerical simulation; the result therefore
motivates improving its optimization rather than replacing it with supervised
training.

Overall, the inverter is difficult to learn because its stiffness is
distributed across several interacting fast and slow modes. This produces
both slowly learned modes within individual residuals and strong imbalance
between residual losses, making the inverter harder to learn than the
modified SMIB despite its smaller stiffness ratio. These findings suggest
future work on architectures with separate representations for fast current
dynamics and slower control and filtering states, together with state-wise
normalization and mode-aware weighting. Developing and validating such
NTK-guided architectures is left for future work, with the goal of replacing
trial-and-error model selection by systematic, physics-aware architecture and
training design.

\section{Conclusion}
This paper applies the Neural Tangent Kernel (NTK) method to assess, interpret, and improve the training of neural surrogate models for power-system dynamics. As physics-informed machine learning is increasingly used to model aggregate power-system behavior, accelerate simulations of complex components such as inverters, and support black-box converter surrogates, understanding how architecture and hyperparameters affect performance becomes essential. We show that classical notions of numerical stability in differential-algebraic systems reappear as convergence, conditioning, and robustness issues in NN training. Through its analogy with small-signal eigenvalue analysis, NTK provides a modal interpretation of learning dynamics, showing how states with widely different timescales are learned and guiding adaptive loss weighting.

Focusing on individual power-system components, we studied physics-informed learning of stiff dynamics using vanilla PINNs, KANs, and ActNets. On \ac{SM} models, KANs and ActNets captured fast and slow states more evenly than vanilla PINNs, with ActNet providing the best accuracy--runtime trade-off. For the 11-state inverter model, the results show that learnability depends less on the overall stiffness range than on the distribution of stiff directions. Although the inverter has a smaller stiffness ratio than the modified SMIB benchmark, its electrical, control, and synchronization dynamics create a highly anisotropic eigenvalue distribution that the NTK inherits. This makes training accurate physics-informed surrogates difficult, and neither NTK-based adaptive weighting nor the tested NN architectures fully resolve the issue. Future work should use NTK-guided architecture design, loss weighting, and optimization strategies to address stiff multi-timescale power-system dynamics and produce fast, accurate neural surrogates.

\bibliographystyle{ieeetr}
\bibliography{references}

\end{document}


\title{Appendix to ``Tools to Explain Neural Networks for Power System Dynamics''}
\author{}
\maketitle
\subsection{Stiffness}
\label{app:stiffness}

Power-system dynamic models are difficult to simulate primarily because they are
\emph{stiff}, meaning that they contain modes evolving on widely separated time scales.
To make this precise, consider the linearized differential dynamics
\[
\dot{\mathbf{x}} = \mathbf{A}\mathbf{x},
\]
where \(\mathbf{A} = \partial f/\partial \mathbf{x}\) is the Jacobian of the
differential part of the model. Let \(\lambda_j\) denote the eigenvalues of
\(\mathbf{A}\).

For stable modes, the decay of each component is governed by \(\Re(\lambda_j)\).
We therefore define the modal decay time scale as
\[
\tau_j = \frac{1}{|\Re(\lambda_j)|}.
\]
If these decay time scales differ by several orders of magnitude, the system is stiff.
A convenient measure is the stiffness ratio
\[
r = \frac{\max_j |\Re(\lambda_j)|}{\min_j |\Re(\lambda_j)|},
\]
where the minimum and maximum are taken over the modes with nonzero real part.
Equivalently,
\[
r = \frac{\tau_{\max}}{\tau_{\min}}.
\]

In power-system component models, stiffness arises because electrical, control,
and electromechanical states evolve on very different time scales. Fast current
or voltage-control modes may decay in sub-milliseconds, while mechanical or
outer-control modes may evolve over tens of milliseconds or even seconds. This
separation is the main reason why accurate time-domain simulation becomes
computationally demanding.

\subsection{Numerical Stability}
\label{app:numerical_stability}

Numerical stability is conveniently understood by applying a time-stepping
scheme to the scalar test equation
\[
\dot u = \lambda u,
\qquad \Re(\lambda)<0.
\]
After one time step of size \(h\), a one-step integration method produces
\begin{equation}
u_{k+1} = R(h\lambda)\,u_k,
\label{eq:stability_time_step}
\end{equation}
where \(R(z)\) is the stability function of the method.

Stability requires
\[
|R(h\lambda)| < 1
\]
for the relevant eigenvalues of the system Jacobian. For explicit Euler,
\[
R(z)=1+z,
\]
so stability requires the time step to resolve the fastest decaying mode:
\[
h \ll \frac{1}{\max_j |\Re(\lambda_j)|}.
\]
Thus, one very fast mode can force a prohibitively small step size even if the
quantities of interest evolve much more slowly.

Implicit methods have much larger stability regions and are therefore preferred
for stiff systems. However, they require the solution of nonlinear algebraic
systems at every time step. In large-scale power-system simulations, this
creates a trade-off: explicit methods are cheap per step but constrained by
stability, whereas implicit methods are stable for larger steps but expensive
because each step requires iterative solves.

This same multiscale structure reappears in the learning problem considered in
this paper. There, stiffness no longer appears as a restriction on the
integration step size, but as an imbalance in the rates at which different
components of the training error decay.

\subsection{Synchronous Generator Models} \label{app:sm_models}

The second-order model, referred to as the \emph{2D scenario}, captures essential rotor dynamics governed by the power balance at the synchronous machine (SM) rotor. The states are the rotor angle~$\delta$ and speed deviation~$\omega$, with dynamics:
\begin{subequations}\label{eq:SM2}
\begin{align}
    \dot{\delta} &= \omega, \\
    \dot{\omega} &= \frac{\omega_b}{2H}\left(P_m - P_e - D\,\omega \right),
\end{align}
\end{subequations}
where $\omega_b$ is the base electrical angular frequency, $H$ is the inertia constant, $P_m$ and $P_e$ are the mechanical and electrical power, respectively, and $D$ is the damping coefficient. Assuming an infinite bus, the electrical power is
$P_e = V_1 V_2 \sin(\delta)$, where $V_1$ is the machine terminal voltage and $V_2$ the infinite-bus voltage magnitude.

\paragraph{4D Scenario.}
The fourth-order model extends the 2D model by including transient internal voltage dynamics $(e_d', e_q')$:
\begin{subequations}\label{eq:SM4}
\begin{align}
    \dot{e_d'} &= \frac{1}{T'_{q0}} \left(-e_d' + i_q (x_q - x_q') \right),\\
    \dot{e_q'} &= \frac{1}{T'_{d0}} \left(-e_q' + i_d (x_d - x_d') + e_{fd} \right),\\
    P_e &= e_d' i_d + e_q' i_q + i_d i_q (x_q' - x_d').
\end{align}
\end{subequations}
Here, $T'_{d0}$ and $T'_{q0}$ are transient open-circuit time constants, $x_{d}, x_{q}$ and $x_d', x_q'$ are synchronous and transient reactances, $i_d,i_q$ are stator currents, and $e_{fd}$ is the field voltage.

\paragraph{6D Scenario.}
The sixth-order model further includes subtransient internal voltage dynamics $(e_d'', e_q'')$:
\begin{subequations}\label{eq:SM6}
\begin{align}
    \dot{e_d''} &= \frac{1}{T''_{q0}}\left(e_d' - e_d'' - i_q (x_q' - x_q'') \right),\\
    \dot{e_q''} &= \frac{1}{T''_{d0}}\left(e_q' - e_q'' - i_d (x_d' - x_d'') \right).
\end{align}
\end{subequations}

\begin{table}[t]
\centering
\caption{Parameters used for synchronous generator modeling.}
\begin{tabular}{ll}
\toprule
\textbf{Parameter} & \textbf{Value} \\
\midrule
\multicolumn{2}{c}{\textbf{Synchronous Machine Parameters}} \\
\midrule
Damping coefficient, $D$ & 2 \\
Inertia constant, $H$ [s] & 5.06 \\
Base angular frequency, $\omega_b$ [rad/s] & 314.159 \\
Transient time constants, $T'_{d0}$, $T'_{q0}$ [s] & 4.75,\ 1.6 \\
Subtransient time constants, $T''_{d0}$, $T''_{q0}$ [s] & 0.03,\ 0.05 \\
Direct-axis reactances, $x_d$, $x_d'$, $x_d''$ [p.u.] & 1.25,\ 0.232,\ 0.20 \\
Quadrature-axis reactances, $x_q$, $x_q'$, $x_q''$ [p.u.] & 1.22,\ 0.715,\ 0.20 \\
Field voltage, $e_{fd}$ [p.u.] & 1 \\
Infinite bus voltage, $V_2$ [p.u.] & 1 \\
\bottomrule
\end{tabular}
\end{table}
\subsubsection{Input Domain}

The input domain $\Omega_d$ defines the space from which initial conditions for the simulations are drawn. Since the present work considers only the synchronous machine dynamics described in the 2D, 4D, and 6D scenarios, the input domain must include initial values for the states involved in these models, namely: rotor angle $\delta$, rotor speed deviation $\omega$, transient voltages $(e_d', e_q')$, and, when applicable, the subtransient voltages $(e_d'', e_q'')$. The sampling ranges for these variables are summarized in Table~\ref{tab:input_domain}. A total of 400 initial conditions were generated via Latin Hypercube Sampling (LHS) over this domain. The same domain and sampling strategy were used to initialize collocation points.

\begin{table}[t]
\centering
\caption{Input-domain variables for synchronous generator simulations.}
\label{tab:input_domain}
\begin{tabular}{ll}
\toprule
\textbf{Variable} & \textbf{Domain} \\
\midrule
\multicolumn{2}{c}{\textbf{Rotor States}} \\
\midrule
Rotor angle, $\delta$ [rad] & $[-0.05,\ 0.05]$ \\
Rotor speed deviation, $\omega$ [rad/s] & $[-0.05,\ 0.05]$ \\
\midrule
\multicolumn{2}{c}{\textbf{Transient and Subtransient Voltages}} \\
\midrule
Transient direct-axis voltage, $e_d'$ [p.u.] & $0$ (fixed) \\
Transient quadrature-axis voltage, $e_q'$ [p.u.] & $[0.99,\ 1.02]$ \\
Subtransient direct-axis voltage, $e_d''$ [p.u.] & $[0.99,\ 1.02]$ \\
Subtransient quadrature-axis voltage, $e_q''$ [p.u.] & $[0.99,\ 1.02]$ \\
\midrule
\multicolumn{2}{c}{\textbf{Mechanical / Field Inputs}} \\
\midrule
Mechanical power, $P_m$ [p.u.] & $0.7048$ (fixed) \\
Field voltage, $e_{fd}$ [p.u.] & $1.08$ (fixed) \\
\bottomrule
\end{tabular}
\end{table}

The rotor angle $\delta$ and speed deviation $\omega$ define the electromechanical initial operating point of the synchronous generator. The transient and subtransient voltages $(e_d', e_q', e_d'', e_q'')$ represent the internal electromagnetic states of the machine and determine short-term voltage dynamics under the 4D and 6D models. Fixed values for the field voltage $e_{fd}$ and mechanical input $P_m$ correspond to steady operating conditions of the generator in the absence of explicit excitation-system or governor dynamics. The selected intervals capture a realistic but sufficiently wide range of perturbations around nominal operating points, enabling robust evaluation of the learned dynamics across multiple machine orders.

\subsection{Second-Order Electromechanical Oscillator with Fast Voltage Dynamics}
\label{app:oscillator}

To study the effect of one dominant fast electrical mode, we consider the
modified SMIB model
\begin{equation}
\begin{bmatrix}
1 & 0 & 0 \\
0 & H & 0 \\
0 & 0 & \tau
\end{bmatrix}
\frac{d}{dt}
\begin{bmatrix}
\delta \\ \omega \\ V
\end{bmatrix}
=
\begin{bmatrix}
\omega \\
P_m - E V \sin(\delta) + P_L(t) - D \omega \\
V_{\mathrm{ref}} - V + k\sin(\delta)
\end{bmatrix}
\label{eq:modified_smib_appendix}
\end{equation}
Equivalently, in explicit form,
\begin{align}
\dot{\delta} &= \omega, \\
\dot{\omega} &= \frac{1}{H}\Bigl(P_m - E V\sin(\delta) + P_L(t) - D\omega\Bigr), \\
\dot{V} &= \frac{1}{\tau}\Bigl(V_{\mathrm{ref}} - V + k\sin(\delta)\Bigr).
\end{align}
Here, $\delta$ denotes the generator rotor angle, $\omega$ is the rotor-speed
deviation, and $V$ is the terminal-voltage magnitude. The parameter $H$
represents the generator inertia, $D$ is the damping coefficient, and $P_m$
is the mechanical power input. The constant $E$ denotes the internal voltage
magnitude, while $P_L(t)$ represents a time-varying power disturbance.
Furthermore, $V_{\mathrm{ref}}$ is the voltage reference, $\tau$ is the
voltage-dynamics time constant, and $k$ determines the coupling between the
rotor angle and the voltage dynamics.

The parameters used in this study are
\[
H = 10~\mathrm{s}, \qquad
D = 1~\mathrm{p.u./(rad/s)}, \qquad
P_m = 1~\mathrm{p.u.},
\]
\[
E = 1~\mathrm{p.u.}, \qquad
\tau = 0.01~\mathrm{s},
\]
with disturbance
\[
P_L(t)=0.5\sin(10t),
\]
and initial conditions
\[
\delta(0)=0, \qquad \omega(0)=0, \qquad V(0)=1.
\]

\paragraph{Local time-scale estimate.}
Because the forcing term \(P_L(t)\) is time-varying, the system is not analyzed
through a single global equilibrium. Instead, the stiffness estimate used in the
paper is obtained from the Jacobian of the \emph{frozen} dynamics around a
nominal operating point \((\delta^\star,\omega^\star,V^\star)\). For the
explicit form above, the Jacobian is
\[
\mathbf{A}(\delta^\star,\omega^\star,V^\star)=
\begin{bmatrix}
0 & 1 & 0 \\
-\dfrac{E V^\star \cos(\delta^\star)}{H} & -\dfrac{D}{H} &
-\dfrac{E\sin(\delta^\star)}{H} \\
\dfrac{k\cos(\delta^\star)}{\tau} & 0 & -\dfrac{1}{\tau}
\end{bmatrix}.
\]

For a nominal operating point with small angle deviation and \(V^\star\approx 1\),
this becomes approximately
\[
\mathbf{A} \approx
\begin{bmatrix}
0 & 1 & 0 \\
-\dfrac{E}{H} & -\dfrac{D}{H} & 0 \\
\dfrac{k}{\tau} & 0 & -\dfrac{1}{\tau}
\end{bmatrix}.
\]

This form shows the key structure directly:
\begin{itemize}
    \item the voltage state contributes a fast electrical mode with decay rate of order \(1/\tau\);
    \item the \((\delta,\omega)\) subsystem contributes the slower electromechanical dynamics, whose decay rate is of order \(D/(2H)\) in the lightly damped regime.
\end{itemize}

Accordingly, the dominant decay rates are approximately
\[
|\Re(\lambda_{\mathrm{fast}})| \approx \frac{1}{\tau} = 100,
\qquad
|\Re(\lambda_{\mathrm{slow}})| \approx \frac{D}{2H} = 0.05.
\]
The corresponding stiffness ratio is therefore
\[
r_{\mathrm{SMIB}}
\approx
\frac{|\Re(\lambda_{\mathrm{fast}})|}{|\Re(\lambda_{\mathrm{slow}})|}
=
\frac{1/\tau}{D/(2H)}
=
\frac{2H}{D\tau}
\approx 2\times 10^3.
\]

This estimate shows that the modified SMIB benchmark is stiff, but in a
comparatively simple way: the stiffness is dominated by essentially one fast
electrical direction, while the remaining dynamics evolve on a much slower
electromechanical time scale.

\subsection{Grid-Following Inverter Model Description}
\label{app:inverter_model}

The grid-following inverter considered in this work combines a phase-locked loop
(PLL), an outer active/reactive power controller, an inner current controller,
an \(RL\) electrical interface, and first-order measurement filters. All
quantities are expressed in the synchronous \(dq\) frame aligned with the PLL angle.
This is a reduced-order PQ-controlled inverter model: it does not explicitly
include a cascaded voltage-control loop, PWM switching, or LC/LCL filter
dynamics. The lower-level voltage-regulation and modulation dynamics are
therefore assumed to be sufficiently fast or abstracted into the commanded
modulation voltages \(v_{md}\) and \(v_{mq}\).

\paragraph{State, input, and parameter vectors.}
The eleven-dimensional state vector is
\[
\mathbf{x} =
\bigl[
\xi_{\mathrm{pll}},\,
\theta_{\mathrm{pll}},\,
i_d,\,
i_q,\,
\xi_{i_d},\,
\xi_{i_q},\,
\xi_P,\,
\xi_Q,\,
P_{\mathrm{filt}},\,
Q_{\mathrm{filt}},\,
V_{\mathrm{filt}}
\bigr]^\mathsf{T}.
\]

The input vector is
\[
\mathbf{u} =
\bigl[
P_{\mathrm{ref}},\,
Q_{\mathrm{ref}},\,
V_{\mathrm{ref}},\,
\omega_{\mathrm{ref}},\,
v_{ga},\,
v_{gb},\,
v_{dc}
\bigr]^\mathsf{T},
\]
and the parameter vector collects the control gains and electrical parameters,
\begin{align*}
\mathbf{p} =
\bigl[
&k_{p_{dp}},\,k_{i_{dp}},\,
k_{p_{dq}},\,k_{i_{dq}},\,
k_{p_{id}},\,k_{i_{id}},\\
&k_{p_{iq}},\,k_{i_{iq}},\,
k_{p_{\mathrm{pll}}},\,k_{i_{\mathrm{pll}}},\,
R,\,L,\,\omega_{\mathrm{filt}}
\bigr]^\mathsf{T}.
\end{align*}

\paragraph{Synchronous-frame voltage transformation.}
The PLL angle defines the synchronous frame:
\begin{align}
v_{gd} &= v_{ga}\cos\theta_{\mathrm{pll}} + v_{gb}\sin\theta_{\mathrm{pll}}, \\
v_{gq} &= -\,v_{ga}\sin\theta_{\mathrm{pll}} + v_{gb}\cos\theta_{\mathrm{pll}}.
\end{align}

\paragraph{Phase-locked loop.}
The PLL regulates the quadrature voltage component \(v_{gq}\) to zero through
\[
\omega_{\mathrm{pll}} =
k_{p_{\mathrm{pll}}}\,v_{gq}
+ k_{i_{\mathrm{pll}}}\,\xi_{\mathrm{pll}},
\qquad
\omega = \omega_{\mathrm{ref}} + \omega_{\mathrm{pll}},
\]
with state equations
\[
\dot{\xi}_{\mathrm{pll}} = v_{gq},
\qquad
\dot{\theta}_{\mathrm{pll}} = \omega.
\]

\paragraph{Outer active and reactive power control.}
The power-tracking errors are
\[
dP = P_{\mathrm{ref}} - P_{\mathrm{filt}},
\qquad
dQ = -Q_{\mathrm{ref}} + Q_{\mathrm{filt}}.
\]
These generate current references through PI controllers:
\begin{align}
i_{d,\mathrm{ref}} &= k_{p_{dp}}\,dP + k_{i_{dp}}\,\xi_P, \\
i_{q,\mathrm{ref}} &= k_{p_{dq}}\,dQ + k_{i_{dq}}\,\xi_Q,
\end{align}
with
\[
\dot{\xi}_P = dP,
\qquad
\dot{\xi}_Q = dQ.
\]

\paragraph{Inner current control and electrical plant.}
The current controllers produce modulation voltages
\begin{align}
v_{md} &= k_{p_{id}}(i_{d,\mathrm{ref}} - i_d)
       + k_{i_{id}}\,\xi_{i_d}
       - \frac{\omega}{\omega_{\mathrm{ref}}}L\,i_q, \\
v_{mq} &= k_{p_{iq}}(i_{q,\mathrm{ref}} - i_q)
       + k_{i_{iq}}\,\xi_{i_q}
       + \frac{\omega}{\omega_{\mathrm{ref}}}L\,i_d,
\end{align}
with integrator dynamics
\[
\dot{\xi}_{i_d} = i_{d,\mathrm{ref}} - i_d,
\qquad
\dot{\xi}_{i_q} = i_{q,\mathrm{ref}} - i_q.
\]

The converter-side \(RL\) network evolves according to
\begin{align}
\dot i_d &= -\frac{R}{L}i_d + \omega i_q + \frac{1}{L}(v_{md} - v_{gd}), \\
\dot i_q &= -\frac{R}{L}i_q - \omega i_d + \frac{1}{L}(v_{mq} - v_{gq}).
\end{align}

\paragraph{Power and measurement filters.}
The instantaneous powers are
\[
P = v_{md}i_d + v_{mq}i_q,
\qquad
Q = v_{mq}i_d - v_{md}i_q,
\]
and the converter-voltage magnitude is
\[
v_m = \sqrt{v_{md}^2 + v_{mq}^2}.
\]
These are filtered through first-order dynamics:
\begin{align}
\dot P_{\mathrm{filt}} &= \omega_{\mathrm{filt}}(P - P_{\mathrm{filt}}), \\
\dot Q_{\mathrm{filt}} &= \omega_{\mathrm{filt}}(Q - Q_{\mathrm{filt}}), \\
\dot V_{\mathrm{filt}} &= \omega_{\mathrm{filt}}(v_m - V_{\mathrm{filt}}).
\end{align}
Here, \(V_{\mathrm{filt}}\) is only a filtered voltage-magnitude measurement and
is not fed back through an explicit voltage-control loop.
Collecting all terms gives the nonlinear state-space model
\[
\dot{\mathbf{x}} = f(\mathbf{x},\mathbf{u},\mathbf{p}).
\]

\paragraph{Nominal operating point.}
Solving \(f(\mathbf{x}^\star,\mathbf{u}^\star,\mathbf{p}^\star)=0\) numerically
yields the nominal equilibrium
\begin{align*}
     \footnotesize \mathbf{x}^\star &\approx  \begin{array}{ccccccc}[-0.3491 & 0& 0.9902 & 0 & 0.00202 & 0& \dots \end{array}\\ &\quad \quad  \quad \begin{array}{ccccc} 0.01980 & 0 & 1.0000 & 0 & 1.0099]^\mathsf{T}\end{array}
\end{align*}

corresponding to
\[
(\xi_{\mathrm{pll}},\theta_{\mathrm{pll}},i_d,i_q,\xi_{i_d},\xi_{i_q},
\xi_P,\xi_Q,P_{\mathrm{filt}},Q_{\mathrm{filt}},V_{\mathrm{filt}}).
\]

\paragraph{Eigenvalue groups and time scales.}
Linearizing the model around \((\mathbf{x}^\star,\mathbf{u}^\star,\mathbf{p}^\star)\)
gives a Jacobian with \(11\) eigenvalues in total. Since complex eigenvalues occur
in conjugate pairs, it is more convenient to report them as \emph{eigenvalue groups}.
Grouping conjugate pairs together, the spectrum is
\begin{align*}
\{&-2552.29,\,-2500.28,\,-622.95,\,-619.97,\,-15.00 \pm 25.98j,\\
  &-25.00,\,-25.01 \pm 25.20j,\,-25.26 \pm 25.05j\}.
\end{align*}

Thus, the table below reports eight eigenvalue groups, which together account
for all \(11\) eigenvalues of the Jacobian.

\begin{table}[t]
\centering
\caption{Dominant eigenvalue groups and associated decay time scales for the linearized inverter model. Complex-conjugate pairs are reported in one row.}
\label{tab:timescales_exact}
\begin{tabular}{m{1cm}m{2cm}m{2cm}m{2cm}}
\toprule
\textbf{Mode Group} & \textbf{Eigenvalue(s)}
& \textbf{Decay time \newline scale [s]}
& \textbf{Dominant \newline Variables} \\
\midrule
1 & $-2552.29$                  & $3.92\times 10^{-4}$ & $i_d,\; i_q,\; \xi_{i_d},\; \xi_{i_q}$ \\
2 & $-2500.28$                  & $4.00\times 10^{-4}$ & $i_d,\; i_q,\; \xi_{i_d},\; \xi_{i_q}$ \\
3 & $-622.95$                   & $1.61\times 10^{-3}$ & $i_d,\; i_q,\; \xi_{i_d},\; \xi_{i_q}$ \\
4 & $-619.97$                   & $1.61\times 10^{-3}$ & $i_d,\; i_q,\; \xi_{i_d},\; \xi_{i_q}$ \\
5 & $-15.00 \pm 25.98j$         & $6.67\times 10^{-2}$ & $\xi_{\mathrm{pll}},\; \theta_{\mathrm{pll}},\; \xi_P,\; \xi_Q$ \\
6 & $-25.00$                    & $4.00\times 10^{-2}$ & $P_{\mathrm{filt}},\; Q_{\mathrm{filt}},\; V_{\mathrm{filt}}$ \\
7 & $-25.01 \pm 25.20j$         & $4.00\times 10^{-2}$ & $P_{\mathrm{filt}},\; Q_{\mathrm{filt}},\; V_{\mathrm{filt}}$ \\
8 & $-25.26 \pm 25.05j$         & $3.96\times 10^{-2}$ & $P_{\mathrm{filt}},\; Q_{\mathrm{filt}},\; V_{\mathrm{filt}}$ \\
\bottomrule
\end{tabular}
\end{table}

\paragraph{Interpretation.}
The inverter exhibits three clearly separated dynamical layers:
\begin{itemize}
    \item \textbf{Fast layer:} the inner current-control and electrical \(RL\) dynamics, with decay time scales between approximately \(4\times 10^{-4}\) s and \(1.6\times 10^{-3}\) s;
    \item \textbf{Intermediate layer:} the measurement filters, with time scales near \(4\times 10^{-2}\) s;
    \item \textbf{Slow layer:} the PLL and outer power-control loop, with a decay time scale near \(6.7\times 10^{-2}\) s.
\end{itemize}

Using the real parts of the fastest and slowest eigenvalues, the corresponding
stiffness ratio is approximately
\[
r_{\mathrm{inv}}
=
\frac{2552.29}{15.00}
\approx 1.7\times 10^2.
\]

Compared with the modified SMIB example, the inverter has a smaller overall
stiffness ratio, but its stiffness is distributed across several interacting
fast directions rather than being dominated by a single fast mode. This
multi-directional stiffness is what makes the inverter substantially harder to
learn in the results section.
